\documentclass[a4paper, 11pt, twoside,openright]{article}

\usepackage[utf8]{inputenc}
\usepackage[T1]{fontenc}
\usepackage[english]{babel}
\usepackage{lipsum}
\usepackage{gensymb}
\usepackage{array}
\usepackage{lmodern}
\usepackage{amsmath}
\usepackage{amssymb}
\usepackage{mathrsfs}
\usepackage{xcolor}
\usepackage{soul}
\usepackage{enumitem}
\usepackage{pifont}
\usepackage{caption}
\usepackage{verbatim}
\usepackage{algorithm}
\usepackage{algorithmic}

\usepackage{setspace}

\usepackage{multirow}
\usepackage{multicol}
\usepackage{longtable}
\usepackage{tabularx}
\usepackage{diagbox}

\usepackage[hidelinks]{hyperref}
\usepackage{fancyref} 
\usepackage[numbers,sort]{natbib}
\usepackage{url}

\usepackage{graphicx}
\graphicspath{{images/}}
\usepackage{float} 
	
\usepackage{titlesec}

\usepackage{fancyhdr}
\usepackage{lineno}

\let\oldpagenumbering\pagenumbering
\renewcommand{\pagenumbering}[1]{%
	\cleardoublepage
	\oldpagenumbering{#1}
}

\newcommand{\vect}[1]{\underline{#1}}
\newcommand{\tens}[1]{\underline{\underline{#1}}}

\newcommand{\refdomain}{\Omega_0}

\newcommand{\posteriorsurfaceref}{\Gamma^{post}_0}

\newcommand{\sclerasurfaceref}{\Gamma^{sclera}_0}

\newcommand{\jacobian}{J}

\newcommand{\Ithree}{I_3}
\newcommand{\Ionebar}{\bar{I}_1}
\newcommand{\Itwobar}{\bar{I}_2}

\newcommand{\virtualpower}{\mathcal{P}}

\newcommand{\stiffnessone}{k_1}
\newcommand{\stiffnesstwo}{k_2}

\newcommand{\stiffnessfibrilapparent}{k_{lamellae,apparent}}

\newcommand{\testfunction}{\vect{w}}
\newcommand{\displVect}{\vect{u}}

\newcommand{\Ftensor}{\tens{F}}
\newcommand{\CGtensor}{\tens{C}}

\newcommand{\GLtensor}{\tens{e}}

\newcommand{\PKstress}{\tens{\Sigma}}

\newcommand{\refconfiggradient}{\tens{\nabla}_{\vect{\xi}}}

\newcommand{\dye}{d_{\vect{u}} \tens{e}}

\newcommand{\energyfunction}{\psi}

\newcommand{\psiIso}{\textcolor{bleufonce}{\psi^{iso}}}
\newcommand{\psiVol}{\textcolor{rose}{\psi^{vol}}}
\newcommand{\psiAniso}{\textcolor{vertjoli}{\psi^{lam}}}

\newcommand{\dpsiAniso}{\textcolor{vertjoli}{\delta \psi^{lam}}}
\newcommand{\dpsiAnisoone}{\textcolor{vertjoli}{\delta \psi^{lam}_{1}}}
\newcommand{\dpsiAnisotwo}{\textcolor{vertjoli}{\delta \psi^{lam}_{2}}}
\newcommand{\dpsiAnisoi}{\textcolor{vertjoli}{\delta \psi^{lam}_{i}}}

\newcommand{\psiisoIoneparameter}{\kappa_1}

\newcommand{\apparentpsiisoIoneparameter}{\kappa_{1}^ {apparent}}
\newcommand{\psiisoItwoparameter}{\kappa_2}

\newcommand{\densityfibone}{\textcolor{vertjoli}{\rho_{1}}}
\newcommand{\densityfibtwo}{\textcolor{vertjoli}{\rho_{2}}}
\newcommand{\inplanefiberdensity}{\kappa_{ip}}
\newcommand{\inplanefiberdensitymin}{\kappa_{ip, min}}
\newcommand{\inplanefiberdensitymax}{\kappa_{ip, max}}

\newcommand{\outofplanefiberdensity}{\kappa_{t}}
\newcommand{\outofplanefiberdensitymin}{\kappa_{t, min}}
\newcommand{\outofplanefiberdensitymax}{\kappa_{t, max}}
\newcommand{\inplanefiberonedensity}{\kappa_{ip,1}}

\newcommand{\inplanefibertwodensity}{\kappa_{ip,2}}

\newcommand{\exfib}{\vect{e}_x^{lam}}
\newcommand{\eyfib}{\vect{e}_y^{lam} }
\newcommand{\ezfib}{\vect{e}_z^{lam} }
\newcommand{\erfib}{\vect{e}_r^{lam} }
\newcommand{\ethetafib}{\vect{e}_\theta^{lam} }
\newcommand{\ephifib}{\vect{e}_\phi^{lam} }

\newcommand{\thetaphi}{(\theta, \phi) }

\newcommand{\rnod}{\vect{r_0}}

\newcommand{\rnodthetaphi}{\vect{r_0}(\theta, \phi)}

\newcommand{\reflenghtfiberone}{l_{0,1}}
\newcommand{\reflenghtfibertwo}{l_{0,2}}
\newcommand{\elongationfiber}{\lambda}
\newcommand{\elongationfiberthetaphi}{\lambda(\theta, \phi)}

\newcommand{\unfoldingelongation}{\lambda_{u}}
\newcommand{\unfoldingelongationmin}{\lambda_{u, min}}
\newcommand{\unfoldingelongationmax}{\lambda_{u, max}}

\newcommand{\unfoldingelongationfibone}{\lambda_{u,1}}
\newcommand{\unfoldingelongationfibtwo}{\lambda_{u,2}}

\newcommand{\tractionnodfibone}{t_{u,1}}
\newcommand{\tractionnodfibtwo}{t_{u,2}}

\newcommand{\rosewriting}[1]{\textcolor{rose}{#1}}

\newcommand{\bleufoncecwriting}[1]{\textcolor{bleufonce}{#1}}

\newcommand{\vertjoliwriting}[1]{\textcolor{vertjoli}{#1}}

\definecolor{rose}{rgb}{1,0,1}
\definecolor{bleuc}{rgb}{0.18,0.46,0.71}
\definecolor{bleufonce}{rgb}{0,0.13,0.38}
\definecolor{vertjoli}{rgb}{0,0.59,0.33}
\definecolor{violetjoli}{rgb}{0.5216,0.0824,0.7804}
\definecolor{purplematlab}{rgb}{0.7344,0.4453, 0.9375}
\definecolor{pinkmatlab}{rgb}{0.8672,0.2773, 0.5547}
\hypersetup{
	colorlinks = true,
	citecolor = rose,
	linkcolor = bleuc
}

\newenvironment{figureth}{%
	\begin{figure}[H]
		\centering
	}{
	\end{figure}
}

\newenvironment{tableth}{%
	\begin{table}[H]
		\centering
	}{
	\end{table}
}

\hypersetup{
	pdfauthor = {Chloe Giraudet},
	pdfsubject = {Paper 1},
	pdftitle = {Patient-specific modeling of the cornea},
	pdfkeywords = {Cornea, biomechanics, multiscale, keratoconus}
}

\begin{document}
\onehalfspacing

	\pagenumbering{roman}

	\pagenumbering{arabic}
	\setcounter{secnumdepth}{6} 
	\title{Multiscale mechanical model based on patient-specific geometry: application to early keratoconus development}
\date{}
\author{C.Giraudet$^{1,2}$, J. Diaz$^{2,1}$, P. Le Tallec$^{1,2}$, J.-M. Allain$^{1,2}$}
\newcommand{\address}[1]{\def\address{#1}}
\address{%
	$^1$ Laboratoire de Mécanique des Solides, CNRS, Ecole Polytechnique, Institut Polytechnique de Paris\\
	$^2$ Inria\\
	
}

\makeatletter
\renewcommand{\maketitle}{%
	\newsavebox{\foo}
	\savebox{\foo}{%
		\begin{minipage}[t]{15cm}%
			\centering\Large\bfseries\@title
			\par\centering\vspace{1em}\normalfont\normalsize\noindent\@author
			\par\vspace{1.5em}\itshape\scriptsize%
			\noindent\centering\begin{tabular}[t]{@{}l}%
				\address
			\end{tabular}
	\end{minipage}}
	\par\vspace{-1.5em}\hfill
	\par\vspace{1em}\noindent\usebox{\foo}
	\par\vspace{0.5em}\noindent\rule{\linewidth}{1pt}}
\makeatother

\maketitle

\renewenvironment{abstract}%
{\par{\noindent\textbf{\abstractname} --- }}
{\par\vspace{.3\baselineskip}\hrule}
\makeatother
\newcommand{\keywords}{\par\vspace{.01\baselineskip}\noindent\textbf{Keywords} --- }

\begin{abstract}
Keratoconus is a pathology of the cornea associated with a tissue thinning and a weakening of its mechanical properties. However, it remains elusive which aspect is the leading cause of the disease. To investigate this question, we combined a multiscale model with a patient-geometry in order to simulate the mechanical response of healthy and pathological corneas under intraocular pressure. The constitutive behavior of the cornea is described through an energy function which takes into account the isotropic matrix of the cornea, the geometric structure of collagen lamellae and the quasi-incompressibility of the tissue. A micro-sphere description is implemented to take into account the typical features of the collagen lamellae as obtained experimentally, namely their orientation, their stiffness and their dispersion, as well as the their unfolding stretch, at which they start to provide a significant force. A set of reference parameters is obtained to fit experimental inflation data of the literature. We show that the most sensitive parameter is the unfolding stretch, as a small variation of this parameter induces a major change in the corneal apex displacement. The keratoconus case is then studied by separating the impact of the geometry and the one of the mechanics. We computed the evolution of the SimK (a clinical indicator of cornea curvature) and elevation maps: we were able to reproduce the reported changes of SimK with pressure only by a mechanical weakening, and not by a change in geomtry. More specifically, the weakening has to target the lamellae and not the matrix. The mechanical weakening leads to elevations close to early stage keratoconus, but our model lacks the remodeling component to couple the change in mechanics with changes in geometry.
\end{abstract}



	\section{Introduction}

Cornea is a critical part of the eye providing two thirds of its optical power through its specific lens shape. In keratoconus disease, the shape of the cornea is progressively altered to become conical, leading to optical aberration and thus to a loss of vision \cite{sedaghat_comparative_2018}. A late detection of the keratoconus imposes a laser surgery with possible complications \cite{vesaluoma_corneal_2000, moilanen_long-term_2003, holzer_femtosecond_2006}. Conversely, if the keratoconus is detected at an early stage, appropriated contact lenses can be used to stop its progression \cite{barnett_contact_2011, downie_contact_2015}. This explains the interest for early diagnosis methods in the literature \cite{cavas-martinez_new_2017}.

Keratoconus origin is not determined as of today: it has been shown to be favored by genetic, but also by mechanical rubbing of the eye \cite{najmi_correlation_2019}. Early keratoconus are associated with both a thinning of the cornea \cite{pinero_corneal_2010} and a decrease of the mechanical properties \cite{ambekar_effect_2011}, combined with a loss of the highly organized structure of the cornea \cite{radner_altered_1998}. However, it is not clear if the thinning is due to the weakening of the cornea or comes first. To tackle this question, we propose a modeling approach in which we can change independently the cornea geometry and its mechanical properties from healthy to keratoconic ones.

Patient-specific images of the cornea are obtained by clinicians using topographers. They give morpho-geometric indicators for an early stage of the keratoconus \cite{pinero_corneal_2010, cavas-martinez_new_2017, cavas-martinez_study_2018}, such as corneal thickness, anterior and posterior surfaces geometries, and pachymetry. On the other hand, cornea mechanical properties are difficult to estimate specifically \textit{in-vivo} \cite{eliasy_determination_2019, kling_corneal_2014}. They have been investigated \textit{ex-vivo} with inflation tests \cite{elsheikh_biomechanical_2008, benoit_simultaneous_2016} or strip stretching \cite{elsheikh_comparative_2005, zeng_comparison_2001}. They show a response similar to other collagen-rich tissues (as aorta \cite{choudhury_local_2009}, tendon \cite{goulam_houssen_monitoring_2011} or skin \cite{lynch_novel_2017}), with a first heel region associated with a low, non-linear, increase of the stress for large stretch, followed by a linear region in which the force increases proportionally to the stretch. Indeed, it has long been known \cite{maurice_structure_1957} that optical and mechanical properties of the cornea are linked to the micro-structural organization of the stroma \cite{ruberti_corneal_2011, meek_corneal_2015}, a collagen-rich tissue made of a plywood of collagen lamellae anchored in a matrix of proteoglycans and keratocytes. It is classically accepted that the mechanical properties arise from a progressive straightening of the lamellae in the heel region, followed by their stretching in the linear part \cite{ashofteh_yazdi_characterization_2020}, as reported for tendon \cite{fang_modelling_2016} for example. Only a few papers have questioned this interpretation, with contradictory observations \cite{benoit_simultaneous_2016, bell_hierarchical_2018}  either due to the probed scales or to the differences in the experimental conditions. 

The techniques used today to image the corneal lamellae are either destructive (as X-rays scattering \cite{newton_circumcorneal_1998, aghamohammadzadeh_x-ray_2004, meek_use_2009}) or with very limited field of view (as transmission electron microscopy \cite{bergmanson_assessment_2005} and scanning electron microscopy \cite{radner_interlacing_2002, feneck_comparative_2018}, which are also destructive, or Second Harmonic Generation microscopy \cite{winkler_nonlinear_2011, latour_vivo_2012, mercatelli_three-dimensional_2017, avila_quantitative_2019}, which is not destructive). The experimental complexity means that the available data are not patient-specific and thus do not represent the variability of the human eyes. 

The organization of the lamellae has been shown to be different in the keratoconic corneas compared to healthy ones \cite{meek_changes_2005, akhtar_ultrastructural_2008}, and so one can expect different mechanical properties. Brillouin microscopy showed that a mechanical loss occurs in the region of the cone in keratoconus \cite{scarcelli_biomechanical_2014, seiler_brillouin_2019}. Still, there is no consensus on the difference of rigidity {\it in-vivo} between healthy and keratoconic corneas \cite{ambekar_effect_2011}. Mechanically, a global difference between healthy and keratoconic cornea has been observed {\it in-vivo} in the change of the diopter under pressure \cite{mcmonnies_corneal_2010}.

Usually, cornea is modeled as an hyperelastic quasi-incompressible material reinforced by fibers \cite{studer_biomechanical_2010, nguyen_inverse_2011, petsche_role_2013, gefen_biomechanical_2009, whitford_biomechanical_2015, montanino_modeling_2018} representing the two families of lamellae. The validation of these models is only done on a few experiments measuring the displacement of the apex (\cite{elsheikh_biomechanical_2008, lombardo_analysis_2014} for human cornea) or the 3D displacement of the anterior surface (\cite{boyce_full-field_2008} in bovine cornea) and exclusively in healthy cases. Note that most models do not include a variation of the mechanical properties through the cornea thickness, while nanoindentation has shown that the anterior part is stiffer than the posterior part \cite{dias_anterior_2013}. 

We propose here a multi-scale and heterogeneous model of the cornea, based on the experimental lamellae orientations. This model is calibrated on the available experimental data, showing the high sensitivity of the response to the pre-strain of lamellae. This model is then implemented in a finite element code to simulate variations of intra-ocular pressure (or bulge test) on patient-specific geometries, thanks to clinical keratometer elevation maps. We show that a mechanical weakening of the cornea is needed to reproduce the reported variation of diopter with pressure \cite{mcmonnies_corneal_2010}, for both healthy and keratoconic geometries. On the other hand, the change in geometry without mechanical variation does not reproduce the keratoconus response. We also show that the mechanical weakening tends to induce a keratoconus shape if we start from a stress-free healthy geometry, but the quasi-incompressibility of the cornea does not allow the thinning observed in keratoconus. All of this point towards the importance in a weakening of the mechanical properties in the development of the keratoconus. Particularly, our analysis shows that a weakening of the collagen lamellae is the most likely to induce the pathology. Our observations support the importance of an early measure of the cornea mechanical response, as well as the importance of treatments strengthening the collagen fibers.
	\section{Methods}
The mechanical problem we solve is an inflation test where the cornea is fixed on a pressure chamber at its border and put under pressure.  A patient-specific mesh is created using clinical elevations and thicknesses maps. The fixation is located at the sclera, the white and very stiff tissue surrounding the cornea. The material response of the cornea is brought by the stroma, modeled as an hyperelastic matrix reinforced by collagen lamellae. The lamellae orientations are extracted from X-rays \cite{aghamohammadzadeh_x-ray_2004} and SHG images \cite{winkler_three-dimensional_2013, petsche_role_2013}.

\subsection{Patient - specific geometry}
To construct a patient-specific mesh, we proceed in two steps. First, we construct an idealized geometry of the cornea using an analytical description: the geometry of the healthy cornea is almost regular and well described by a parametric quadratic equation \cite{gatinel_corneal_2011}. Considering the apex of the cornea at the origin of a coordinate system with the z-axis oriented vertically and downwards, the anterior and posterior surfaces of the cornea are described by the biconic function \cite{janunts_parametric_2015}: 
\begin{equation}
z(x, y, R_x, R_y, Q_x, Q_y) = z_0 + \frac{ \displaystyle \frac{x^2}{R_x}+\frac{y^2}{R_y} }{\displaystyle 1 + \sqrt{1  - (1 + Q_x) \frac{x^2}{R_x^2}  - (1 + Q_y) \frac{y^2}{R_y^2} } } ,
\label{biconic_equation}
\end{equation}
where $R_x $ and $R_y$ are the curvature radii of the flattest ($x$ axis) and the steepest ($y$ axis) meridians of the cornea, $Q_x$ and $Q_y$ are the associated asphericities. Note that the $x$ and $y$ directions can be rotated of an angle $\psi$ from the classical nasal-temporal (N-T) and inferior-superior (I-S) axes (see Fig.~\ref{corrected_mesh}b for illustration of the anterior surface). Finally, $z_0$ is the arbitrary translation with respect to the $z$ axis origin. 

To adapt the mesh to real cornea, we use anonymized clinical data obtained by an anterior segment OCT combined with a MS-39 placido type topographer (Dr. J. Knoeri's personal communication). Figures~\ref{cornea_mesh_OCT}a, c, g and i present the maps of clinical anterior and posterior elevations for a healthy (Fig.~\ref{cornea_mesh_OCT}a and c) and a keratoconic cornea (Fig.~\ref{cornea_mesh_OCT}g and i). For each surface, a best fit sphere (BFS) is determined during the acquisition. The distance between the BFS and the real surfaces are called the anterior and posterior elevations (for the exterior and interior surface of the cornea respectively). Figures~\ref{cornea_mesh_OCT}e and k show clinical maps of the thicknesses of the same cornea. We first do a least square minimization of Eq.~\eqref{biconic_equation} with respect to the clinical data. Then, the cornea's thickness at the apex is used to place the anterior surface with respect to the posterior surface. This is used to create an idealized mesh (see Fig.~\ref{corrected_mesh}a - grey mesh) thanks to the code provided by Pr. A. Pandolfi \cite{pandolfi_model_2006}.

This mesh is then corrected to match the real one. First, we adjust the anterior and posterior surfaces to match exactly the clinical observations (see Fig.~\ref{corrected_mesh}a - pink mesh). This step requires the interpolation of the elevation maps at the node positions, which is done with a bi-dimensional B-spline approximation. Second, the points in the volume of the mesh (so between the interior and exterior surfaces) are corrected to be linearly distributed between the two surfaces. This procedure ensures that the mesh is both realistic and regular. 

At the end of the process, elevations (Fig.~\ref{cornea_mesh_OCT}b and d for healthy cornea and Fig.~\ref{cornea_mesh_OCT}h and j for keratoconic cornea) and thicknesses (Fig.~\ref{cornea_mesh_OCT}f for healthy cornea and Fig.~\ref{cornea_mesh_OCT}l for keratoconic cornea) are reproduced on the mesh to be compared to the clinical ones. Although they are determined at different positions and thus cannot be compared directly, we can say that the B-splines approximation captures the clinical data (elevations and thicknesses) pretty well, despite the expected tendency to smooth the shape.

An important point is that this mesh is built in the loaded configuration where the cornea is subjected to the physiological intra-ocular pressure (IOP). We call this configuration $\Omega_{physio}$. 

\begin{figureth}
	\includegraphics[width = 1.0\linewidth]{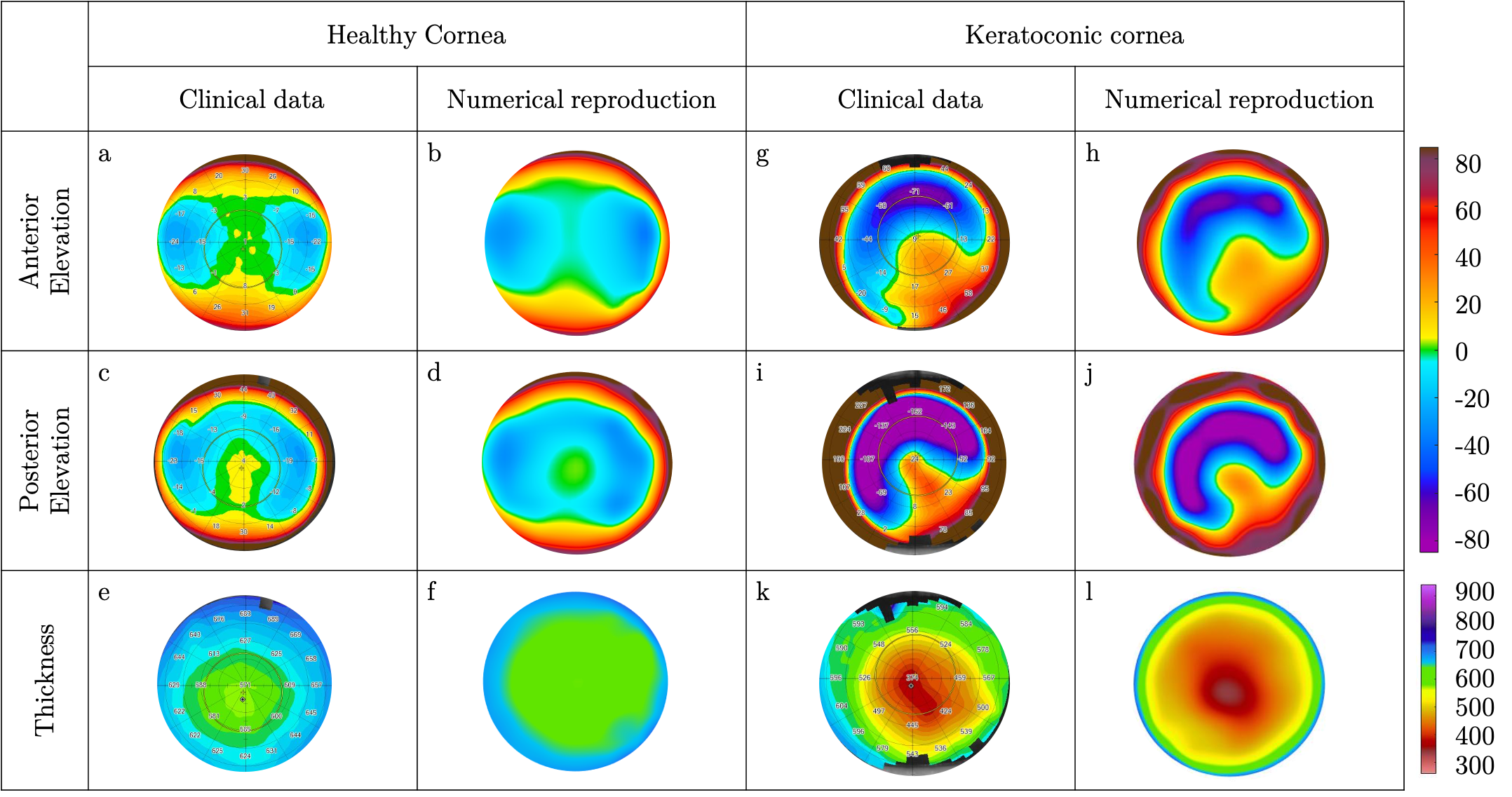}
	\caption[Elevations and thicknesses clinical and computed maps]{\label{cornea_mesh_OCT}Elevation and thickness maps of healthy and keratoconic cornea. (a-f) Clinical and computed maps for a healthy cornea. (g-l) Clinical and computed maps for an advanced stage of keratoconic cornea. (a, c, e, g, i, k) Clinical data obtained by an OCT combined with a MS-39 placido type topographer. (b, d, f, h, j, l) Computed maps at physiological pressure for the same corneas and adapted meshes. (a, b, g, h) Clinical and computed anterior elevations with respect to the best fit sphere (BFS). Scale bar in $\mu  m$. (c, d, i, j) Clinical and computed posterior elevations with respect to the BFS. Scale bar in $\mu  m$. (e, f, k,l) Clinical and computed thickness. Scale bar in $\mu m$.}
\end{figureth}

\begin{figureth}
	\includegraphics[width = 0.7\linewidth]{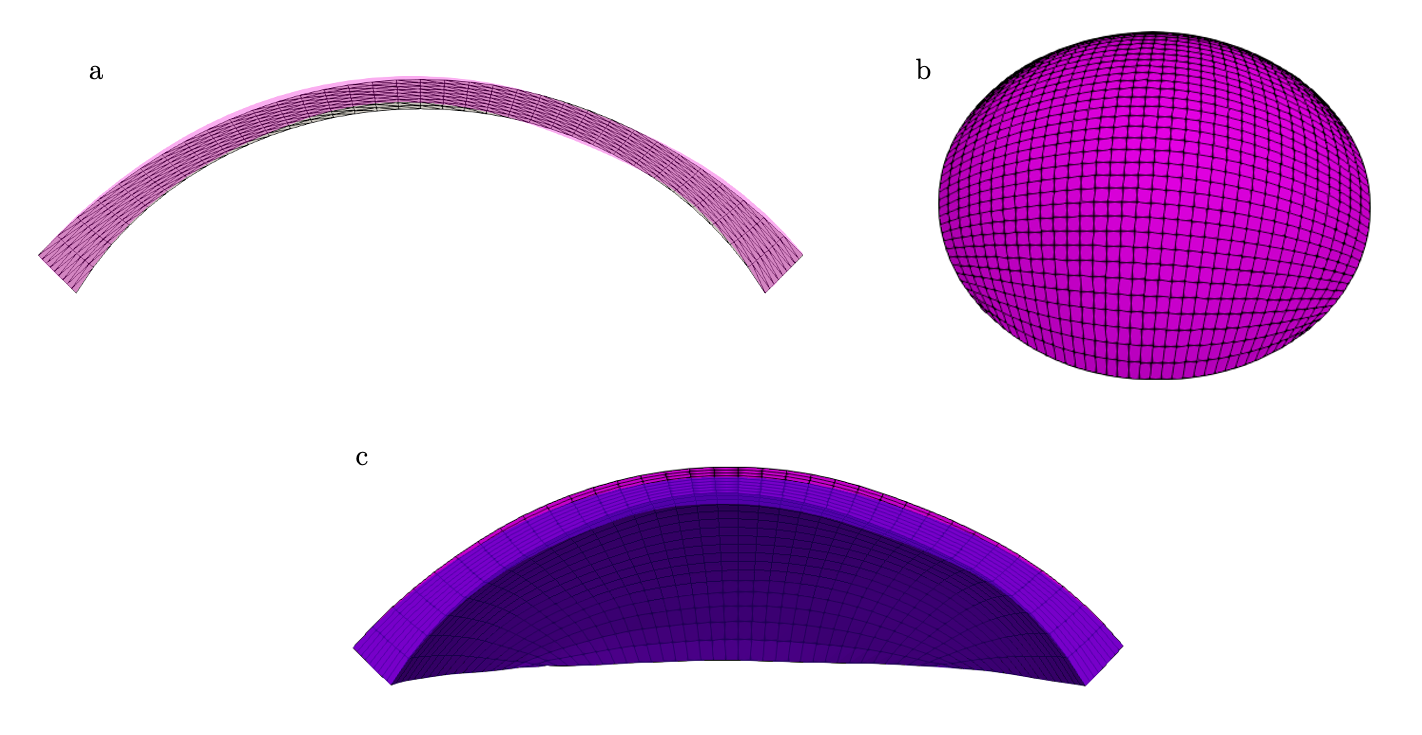}
	\caption[Mesh used in the finite element code process]{\label{corrected_mesh}Example of a mesh construction for a keratoconic cornea. Mesh parameters: 12250 nodes and 10404 hexahedral elements. (a) Vertical cross-section along the long axis of the cornea of the idealized mesh (grey) and the patient-specific mesh at physiological pressure $\protect \Omega_{physio}$ (pink). (b) 3D picture of the patient-specific mesh at physiological pressure $\protect \Omega_{physio}$. (c) Cross-section through the apex of the patient-specific mesh at physiological pressure  $\protect \Omega_{physio}$ (pink) and in stress-free configuration $\protect \Omega_{0, stress-free}$ (blue) to be defined later.}
\end{figureth} 

\subsection{Mechanical equilibrium of the cornea: variational formulation \label{mechanical_equilibrium_cornea}}

We use a weak formulation written in the unknown unloaded configuration $\refdomain$ to represent the energetic equilibrium, the different terms being summarized in Fig.~\ref{inflation_test_meca}. This writes:
\begin{equation}
\displaystyle \virtualpower_{i} = \virtualpower_{e} + \virtualpower_{sclera},
\end{equation}
where $\virtualpower_{i}$ is the inner power, $\virtualpower_{e}$ is the power of external forces and $\virtualpower_{sclera}$ is the power associated to the elastic boundary conditions. We look for a quasi-static solution of the problem, where the inertia terms are neglected. We also neglect volumic forces. The external forces are associated to the pressure $P$ applied on the posterior surface of the cornea, producing a virtual power in Lagrangian formalism: 
\begin{equation}
	\displaystyle \forall \testfunction \in  \mathcal{V}(\refdomain) \text{,} \quad  \virtualpower_{e} =  - P \int_{\posteriorsurfaceref} \jacobian \vect{n}_0 .\Ftensor^{-1} .\testfunction \text{d} \Gamma, 
\end{equation}
with $\testfunction$ an admissible test function (satisfying the boundary conditions), $J = \det (\Ftensor)$ the change in volume, $\Ftensor$ the gradient of the transformation sending $\refdomain$ to $\Omega (t)$ and $\vect{n}_0$ the external normal on the posterior surface in the stress-free configuration. 
The anterior surface is free of loading. The stiff sclera fixed to the pressure chamber is treated as an elastic support boundary condition, producing the virtual power:
\begin{equation}
	\displaystyle \forall \testfunction \in  \mathcal{V}(\refdomain) \text{,} \quad  \virtualpower_{sclera} =  - \int_{\sclerasurfaceref} a \displVect. \testfunction \text{d} \Gamma, 
\end{equation}
 with $\displVect$ the displacement vector, and $a$ the boundary elastic modulus, assumed to be large with respect to the cornea stiffness.

\begin{figureth}
	\includegraphics[width = 1.0\linewidth]{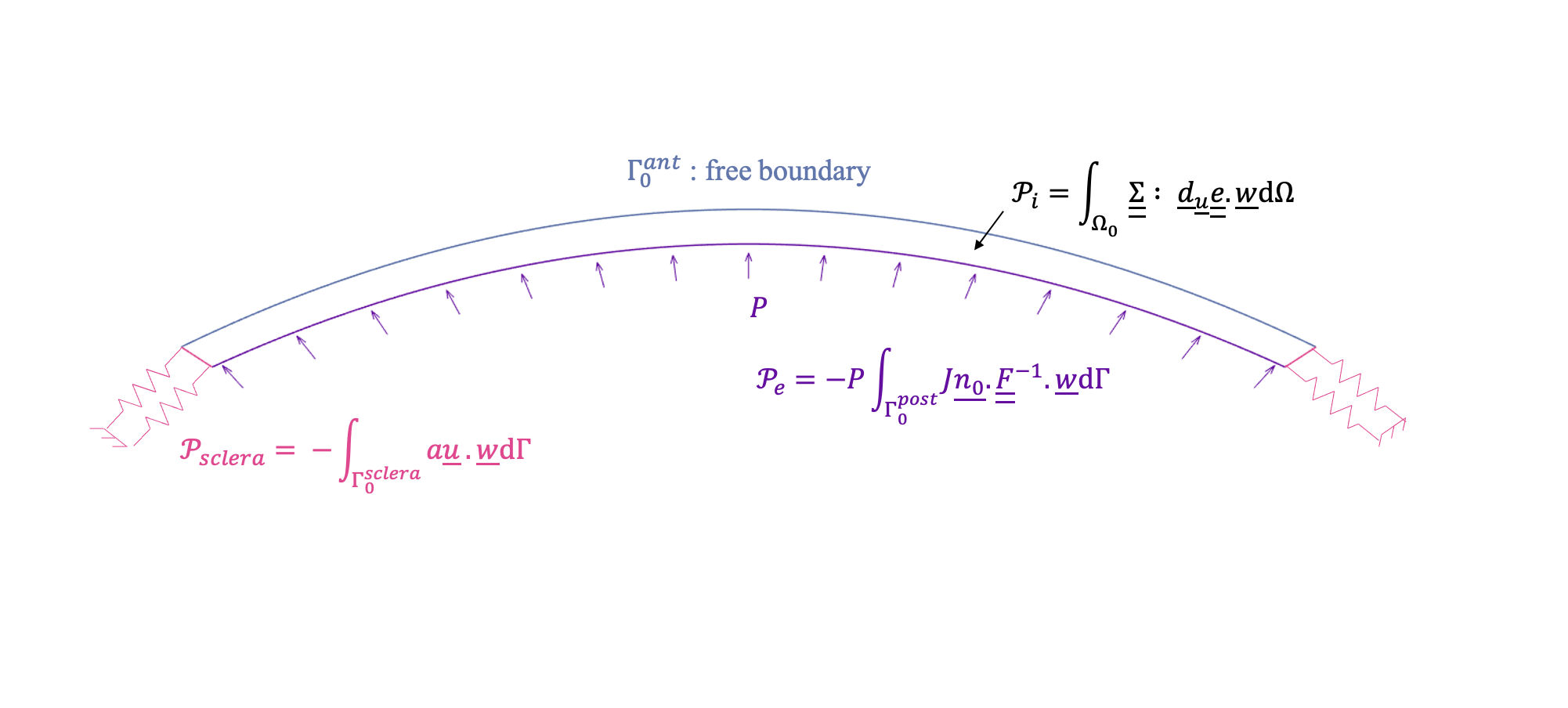}
	\caption[Schematic view of the mechanical problem of the inflation test.]{\label{inflation_test_meca} Schematic view of the mechanical problem of an inflation test. A pressure $P$ is applied on the posterior surface of the cornea, while the anterior surface of the cornea is stress-free, and the sclera is fixed to a pressure chamber treated as an elastic boundary condition of stiffness $a$.}
\end{figureth} 	

Finally, the internal power:
\begin{equation}
\displaystyle \forall \testfunction \in  \mathcal{V}(\refdomain) \text{,} \quad \virtualpower_{i} = \int_{\refdomain} \PKstress : \dye.\testfunction \text{d} \Omega,
\end{equation}
introduces the $2^{nd}$ Piola-Kirchhoff stress tensor $\PKstress$, which is related to the energy function $\energyfunction$ through its derivative with respect to the Green-Lagrange tensor $\displaystyle \GLtensor = \frac{1}{2} (\Ftensor^{T}\Ftensor - \tens{1})$:
\begin{equation}
	\PKstress :=  \frac{d \psi}{d \GLtensor},
\end{equation} 
and $ \dye. \testfunction= \frac{1}{2} (( \refconfiggradient \testfunction) ^T . \Ftensor +\Ftensor^T. \refconfiggradient \testfunction) $, the symmetric part of the gradient tensor of the test function in the current configuration brought back in the reference configuration.\\

The weak formulation of our mechanical problem leads to the following equilibrium equation in Lagrangian form:
\begin{equation}
\forall \testfunction \in \mathcal{V}(\refdomain) \text{,} \quad \int_{\refdomain} \PKstress : \dye.\testfunction \text{d} \Omega = - P \int_{\posteriorsurfaceref} \jacobian \vect{n}_0 .\Ftensor^{-1} . \testfunction  \text{d} \Gamma - \int_{\sclerasurfaceref} a \displVect . \testfunction \text{d} \Gamma .
\label{final_weak_formulation_Sigma}
\end{equation}

\subsection{Constitutive behavior \label{constitutive_behavior_cornea}}

We consider that the mechanical resistance of the cornea arises from the stroma, its main layer \cite{pandolfi_fiber_2012, simonini_customized_2015}. The stroma is a collagen-rich tissue that we describe as a hyperelastic material made of fibers in an isotropic matrix viewed as weakly compressible. So, our associated energy function $\energyfunction$ is splitted into three contributions:
\begin{equation}
\energyfunction = \psiIso + \psiVol + \psiAniso,
\label{decomposition_Psi}
\end{equation}
with an isotropic part $\psiIso$  corresponding to the matrix, the keratocytes and the randomly distributed lamellae, a volumetric part $\psiVol$ penalizing any change of volume and an anisotropic part $\psiAniso$, taking into account the mechanical role of the oriented lamellae. 

The isotropic part of the function $\psiIso$ is chosen here as a Mooney-Rivlin function of the reduced invariants $\Ionebar = I_1 I_3^{-1/3}$ and $\Itwobar = I_2 I_3^{-2/3}$ \cite{pandolfi_model_2006, simonini_customized_2015} of the Cauchy-green tensor  $\CGtensor = \Ftensor^T \Ftensor$: 
\begin{equation}
\psiIso \bleufoncecwriting{:= \psiisoIoneparameter (\Ionebar-3) + \psiisoItwoparameter(\Itwobar-3) } ,
\label{psi_iso}
\end{equation}
while the volumetric part $\psiVol$ penalizes any volumic change by a very large bulk modulus \rosewriting{\textit{K}} \cite{simonini_customized_2015}
\begin{equation}
\psiVol \rosewriting{:= \displaystyle K(J^2 - 1 - 2log(J)),} \quad \text{with} \quad J^2 = \Ithree .
\label{psi_vol}
\end{equation}

The anisotropic contribution is due to the anisotropic distribution of the lamellae. X-ray and SHG observations have shown a two-peak distribution of lamellae (see Fig.~\ref{MeekData}) \cite{aghamohammadzadeh_x-ray_2004, latour_vivo_2012} that we describe by two families of lamellae $\vertjoliwriting{ (lam_1, lam_2) }$. We model their contribution by an angular integration (AI) approach \cite{studer_biomechanical_2010, petsche_role_2013}. At each material point of the cornea, the two families of lamellae have a given directional density distribution $\vertjoliwriting{ (\densityfibone \thetaphi, \densityfibtwo \thetaphi) }$. The contribution $\psiAniso$ of the two families of lamellae at each point adds local contributions of all possible directions, through the integration on a sphere of radius 1 (called "micro-sphere"):
\begin{equation}
\psiAniso \vertjoliwriting{ := \int_{\theta = 0}^{\pi} \int_{\phi = 0}^{2 \pi}{ (\densityfibone \thetaphi \dpsiAnisoone \thetaphi + \densityfibtwo \thetaphi \dpsiAnisotwo \thetaphi) \sin \theta \text{d} \theta  \text{d} \phi} }
\label{psi_aniso_continuous}
\end{equation}
performed in the local system of coordinates at the given spatial quadrature point $(\erfib, \ethetafib, \ephifib)$ (see Fig.~\ref{MeekData}). At each mesh node, a local Cartesian basis $(\exfib, \eyfib, \ezfib)$  (see Fig.~\ref{MeekData}d) is created using the main directions of the lamellae extracted from \cite{aghamohammadzadeh_x-ray_2004}: $\exfib$ is in the direction of one lamellae (chosen as the one which direction is closer to the long axis of the cornea in the central part and the one closer to the tangential direction in the periphery) interpolated at the node from the data at the experimental points; $\ezfib$ is normal to the surface at the node and $\eyfib$ completes the trihedron. Then, $(\erfib, \ethetafib, \ephifib)$ define the local spherical system characterizing the direction $(\theta, \phi)$ of a particular quadrature point of the micro-sphere (see Fig.~\ref{MeekData}e). 

\begin{figureth}
	\includegraphics[width = 0.9\linewidth]{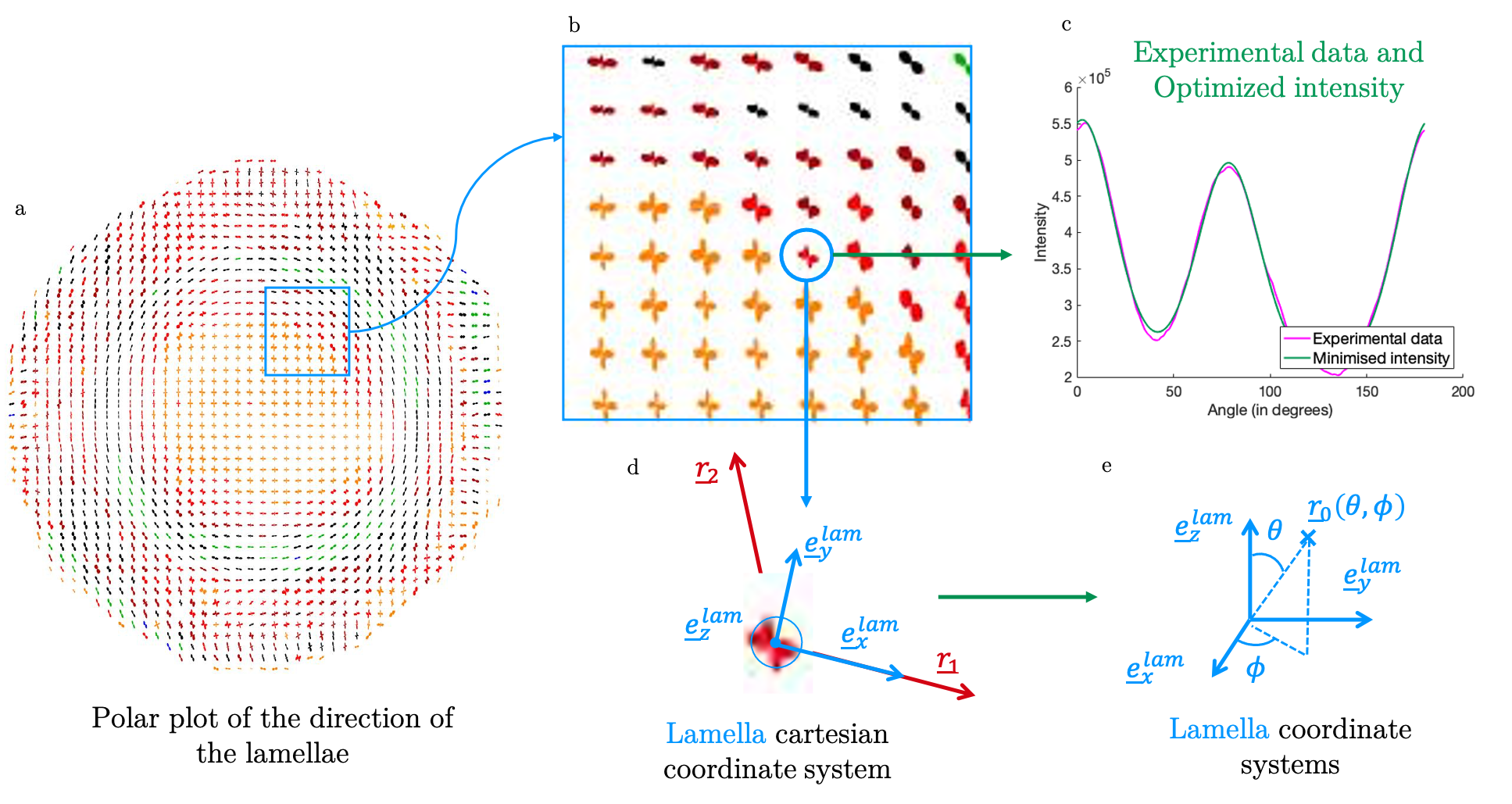}
	\caption[Meek's data and microsphere.]{\label{MeekData} Distribution of lamellae orientation in a cornea. (a) Experimental polar plot of the direction of the lamellae obtained from X-ray observation (Figure from \cite{aghamohammadzadeh_x-ray_2004}, kindly provided by S. Hayes and  K. M. Meek). (b) Zoom on a sub-region of the cornea. (c) Experimental (pink) and associated optimized angular intensity (green) at one point of measurement. (d-e) Local Cartesian coordinates system $  \displaystyle ( \protect \exfib, \protect \eyfib, \protect \ezfib)$ at the same particular point of measurement, and the associated spherical coordinates.}
\end{figureth} 	

\subsubsection{Elementary response of a lamella $\dpsiAniso$ }

In many tissue, collagen fibrils are crimped \cite{fratzl_collagen_2008}, explaining the non-linear response of the tissue, with a heel-region in which the crimps disappear, generating a low force, and a linear region where the fibrils are stretched (and aligned) with a spring-like behavior. In cornea, the collagen fibrils appear very aligned in lamellae \cite{winkler_nonlinear_2011}. Still, they can buckle, but we expect that this buckling occurs at a stretch smaller than the one at physiological pressure. Note that experiments on cornea strips have shown that the fibrils are tilted and that this tilt decreases in the heel region to create the non-linear response, as the crimps in other tissues \cite{bell_hierarchical_2018}.

\begin{figureth}
	\includegraphics[width=10cm]{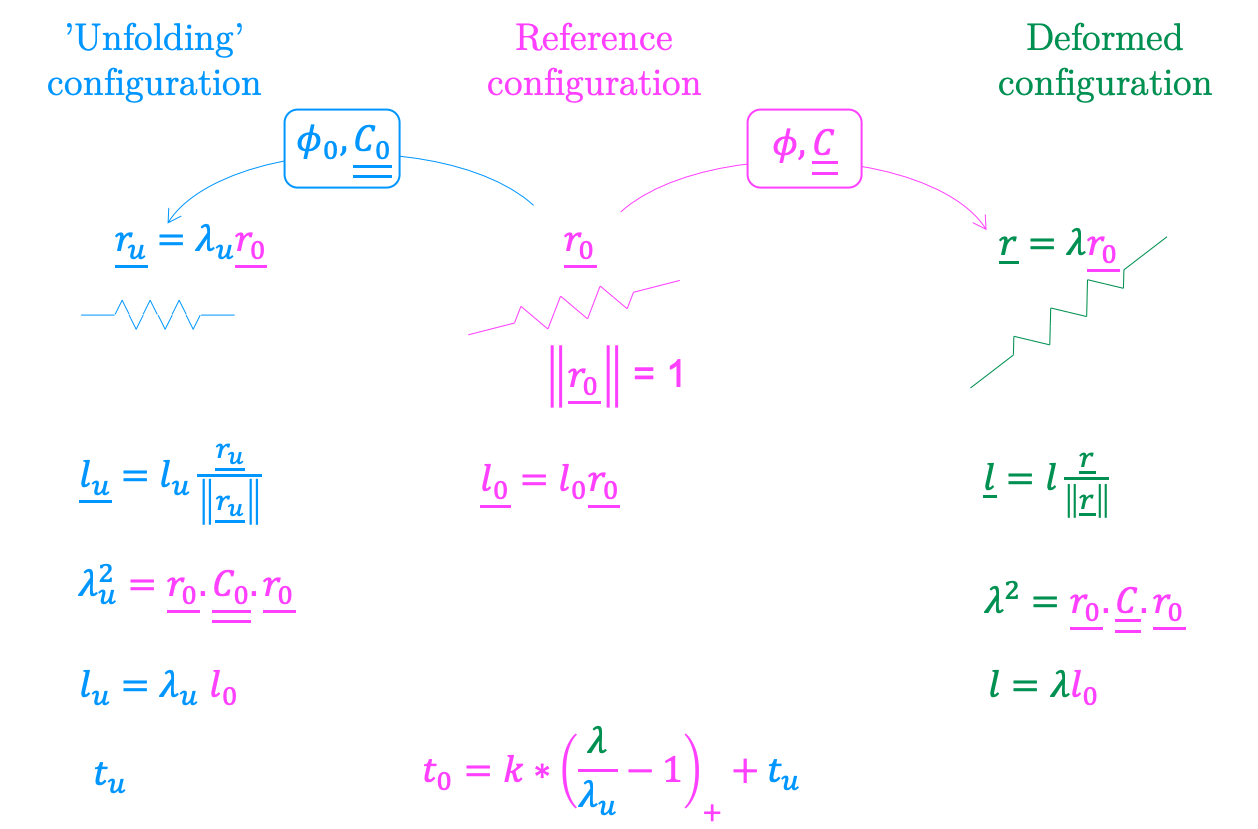}
	\caption[Schematic representation of the different configurations of the lamella]{\label{fiber_configuration} Schematic representation of the different configurations of the lamella: the 'unfolding' configuration corresponds to the limit of the lamella in compression, the reference and deformed configuration are those considered in our problem.}
\end{figureth}

We model a collagen lamella as a bi-domain material (see Fig.~\ref{fiber_configuration}). For stretches below an "unfolding" stretch $\lambda_{u}$, the lamella creates a constant prestress $t_u$, while for higher stretches, the lamella has a spring-like behavior of apparent "stiffness" $k$. The elementary energy function is therefore given by:

\begin{equation}
\displaystyle  \dpsiAnisoi \thetaphi := \frac{1}{2} k_i \lambda_{u , i}  l_{0 , i}( \frac{ \elongationfiber_i}{\lambda_{u , i}  } - 1) _{+}^2 + t_{u , i} l_{0 , i} \elongationfiber_{i}, \quad \forall i \in [1:2],
\label{dpsianiso_without_compression}
\end{equation}
where $()_+$ is the positive absolute value function.

The elongation $\elongationfiber \thetaphi$ of a lamella of reference direction $\rnod \thetaphi$ is directly obtained under an affine assumption as a function of the Cauchy Green tensor: 
\begin{equation}
\elongationfiberthetaphi := \displaystyle \sqrt{ \frac{\rnodthetaphi.\CGtensor.\rnodthetaphi}{\rnodthetaphi.\rnodthetaphi} } = \sqrt{ \rnodthetaphi.\CGtensor.\rnodthetaphi} \quad (||\rnod||^2 = 1), 
\label{stretch_computation}
\end{equation}
with $\rnodthetaphi := \sin \theta \cos \phi \exfib + \sin \theta \sin \phi \eyfib + \cos \theta \ezfib$.

\subsubsection{Density functions $\vertjoliwriting{ (\densityfibone \thetaphi, \densityfibtwo \thetaphi) }$}

The distribution of each lamellae family is described by a Von Mises distribution (Eq.~\eqref{VonMises_distribution}):
\begin{equation}
\text{VM} (\theta, \phi | \inplanefiberdensity, \outofplanefiberdensity, \mu, \nu) :=  \frac{\displaystyle e^{ \inplanefiberdensity \cos(2(\phi-\mu)) } e^{ \outofplanefiberdensity \cos(2(\theta-\nu)) }}{\displaystyle C_{lam} },
\label{VonMises_distribution}
\end{equation}
where $C_{lam}$ is a normalization factor ensuring that the distribution has a total density over the sphere equal to 1. The in-plane $\inplanefiberdensity$ and out-of-plane $ \outofplanefiberdensity$ concentrations are a measure of the dispersion (the larger the $\kappa$ the thinner the peak) when $\mu$ and $\nu$ describe the mean orientations (in-plane and out-of-plane respectively).

To reproduce the X-ray experimental data from \cite{aghamohammadzadeh_x-ray_2004} at each point of measure (see Fig.~\ref{MeekData}c), we consider that the diffracted signal is the sum of the two in-plane distributions of the lamellae families, supplemented by an isotropic contribution:
\begin{equation}
I_{m} (\phi | \inplanefiberonedensity, \inplanefibertwodensity, \mu_1, \mu_2) = I_{iso} + C_1 \text{VM}_{ip} (\phi | \inplanefiberonedensity, \mu_1) + C_2\text{VM}_{ip} (\phi |  \inplanefibertwodensity, \mu_2) ,
\label{I_minimised}
\end{equation}
where $ I^{iso} $ is a constant component representing the isotropic part of the measure, $\mu_1 + \pi/2$ and  $\mu_2 + \pi/2$ the mean directions of the lamellae (the intensity pic is shifted of $\pi/2$ with respect to the main direction of the lamellae \cite{aghamohammadzadeh_x-ray_2004}), $ \inplanefiberonedensity$ and $\inplanefibertwodensity$ the concentrations of the lamellae distributions, and $C_1$ and $C_2$ the measures of the number of oriented lamellae in each direction at the point of measurement. The seven fields $C_1$, $C_2$, $\inplanefiberonedensity$, $\inplanefibertwodensity$, $\mu_1$, $\mu_2$ and $ I_{iso} $, identified at those experimental points by a least square minimization technique, are then bi-linearly interpolated at each node of the mesh.

The X-rays experiments do not give any indication on the out-of-plane distribution. Using Second Harmonic Generation (SHG), it has been shown that the lamellae have a maximum out-of-plane angle of around 30$\degree$ for healthy cornea in the anterior region, well represented by a Gaussian distribution \cite{winkler_three-dimensional_2013} and that the maximum out-of-plane angle decreases with the depth \cite{petsche_role_2013, winkler_three-dimensional_2013}. So, we assumed that the out-of-plane Von Mises distribution has a in-plane mean orientation ($\nu = 0$) so that it reduces to $\displaystyle \text{VM}_t (\theta | \outofplanefiberdensity) =  \frac{e^{ \outofplanefiberdensity \cos(2\theta) }}{\displaystyle C(\outofplanefiberdensity) }$, and that the out-of-plane concentration varies exponentially with depth \cite{petsche_role_2013}: 
\begin{equation}
\displaystyle \outofplanefiberdensity(s) = (\outofplanefiberdensitymin - \outofplanefiberdensitymax) * \frac{ e^{\gamma(1-s) }-1 }{e^{\gamma } -1} + \outofplanefiberdensitymax,
	\quad \text{with} 
	\begin{cases}

	\gamma = 3.19,\\
	\outofplanefiberdensitymin = 7, \\
	\outofplanefiberdensitymax = 700,
	\end{cases}
	\label{cutoffangle_outofplanedensity}
\end{equation}
where $s$ is the normalized depth ($0$ at the anterior surface, $1$ at the posterior), and $\displaystyle C(\outofplanefiberdensity)$ normalizes the distribution. $\outofplanefiberdensitymin $ and $\outofplanefiberdensitymax$ have been chosen such that the maximum cut-off-angle is around $30 \degree$ on the anterior surface ($\outofplanefiberdensity = \outofplanefiberdensitymin$ and so the peak of the distribution is large) and around $0 \degree$ (in-plane lamellae) on the posterior surface of the cornea ($\outofplanefiberdensity = \outofplanefiberdensitymax$ and so the peak of the distribution is tight). No lateral heterogeneity in the lamellae out-of-plane distribution has been reported. 

\subsection{Parameters of the mechanical model}

Once the lamellae orientations are known, our model has still $11$ parameters to be determined: $2$ for the isotropic energy $\psiIso$ ($\bleufoncecwriting{\psiisoIoneparameter} $ and $\bleufoncecwriting{\psiisoItwoparameter}$), $1$ for the volumic energy $\psiVol$ ($\rosewriting{\displaystyle K}$) and $8$ for the anisotropic energy $\psiAniso$ ($\vertjoliwriting{k_i}$, $\vertjoliwriting{\lambda_{u,i}}$, $\vertjoliwriting{ l_{0,i}}$ and $\vertjoliwriting{t_{u,i}}$). Furthermore, all of them except $\rosewriting{\displaystyle K}$ have to be distributed locally to represent the variation of the micro-structure of the cornea. 

The isotropic energy function $\psiIso$ (Eq.~\eqref{psi_iso}) involves two parameters: $\bleufoncecwriting{\psiisoIoneparameter}$ and $\bleufoncecwriting{\psiisoItwoparameter}$. For simplicity, as we have no specific information, we are going to assume that they are proportional with each other:
\begin{equation}
\bleufoncecwriting{\psiisoItwoparameter} = \displaystyle \alpha \bleufoncecwriting{\psiisoIoneparameter}.
\label{Iiso2}
\end{equation}
with $\alpha$ a constant to be identified. We will also make the assumption that they are proportional to the fraction of the isotropic part of the signal $I_{iso}$ (Eq.~\eqref{I_minimised}), so they are distributed in space:
\begin{equation}
\bleufoncecwriting{\psiisoIoneparameter(x,y,s) = \apparentpsiisoIoneparameter} * I_{iso}(x,y,s).
\label{kappa_1_apparent}
\end{equation}
 We consider that this term varies in the cornea's thickness, since the elastic modulus of the posterior stroma is reported to be $39.3\%$ of the modulus of the anterior stroma \cite{dias_anterior_2013}. We thus apply the same exponential variation as for the out-of-plane angular distribution (Eq.~\eqref{kappa1throughthickness}), namely:
\begin{center}
	$
\displaystyle  I_{iso}(x,y,s)  = ( I_{iso}^{ant}(x,y) - I_{iso}^{post}(x,y) ) * \frac{ e^{\gamma(1-s) }-1 }{e^{\gamma } -1} + I_{iso}^{post}(x,y)  ,
$
\end{center}
\begin{equation}
\quad \text{with} 
\begin{cases}
\gamma = 3.19,\\
I_{iso}^{ant}(x,y) \quad \text{depending of the in-plane position} (x,y)  \\
I_{iso}^{post}(x,y) = 39.3\%  * Int_{iso}^{ant}(x,y) .
\end{cases}
\label{kappa1throughthickness}
\end{equation}
Here $I_{iso}^{ant}$ is being obtained by equaling the mean of $I_{iso}(x,y,s)$ in $s$ with the experimental value $I_{iso}$ obtained from the X-ray data. In the end, only $\apparentpsiisoIoneparameter$, a global parameter, needs to be determined to reproduce the experimental data. 

The volumetric energy function $\psiVol$ (Eq.~\eqref{psi_vol}) involves an independent penalty parameter $\rosewriting{K}$ to impose volume conservation, which we consider as a global constant parameter, and which needs to be determined through experimental data.

The anisotropic energy functions $\dpsiAnisoone$ and $\dpsiAnisotwo$ of the two lamellae families (Eq.~\eqref{dpsianiso_without_compression}) involve eight local parameters: $\stiffnessone, \unfoldingelongationfibone, \reflenghtfiberone, \tractionnodfibone, \stiffnesstwo, \unfoldingelongationfibtwo, \reflenghtfibertwo$ and $\tractionnodfibtwo$ (four per lamellae family).\\
$\tractionnodfibone$ and $ \tractionnodfibtwo$ are the forces generated by "undulated" lamellae, which are much smaller than the ones of the stretched ones. So, we are going to neglect them for simplicity, taking $ \tractionnodfibone  = \tractionnodfibtwo = 0$. Thus, the energy functions (Eq.~\eqref{dpsianiso_without_compression}) reduce to: 
$$\displaystyle \dpsiAnisoi  \thetaphi := \frac{1}{2} k_i \lambda_{u,i}  l_{0,i} ( \frac{ \elongationfiber_i}{\lambda_{u,i}  } - 1) _{+}^2, \quad \forall i \in [1:2].$$
The product $\lambda_{u,i} l_{0,i}$ of the unfolding elongation and reference length is the unfolding length of a lamellae $l_{u,i}$. We are assuming that all the lamellae are the same and thus have the same unfolding length: $l_{u,1} = l_{u,2} = l_{u} = Cte$. So the energy function becomes $$\displaystyle \dpsiAnisoi  \thetaphi := \frac{1}{2} k_i l_u ( \frac{ \elongationfiber_i}{\lambda_{u,i}  } - 1) _{+}^2, \quad \forall i \in [1:2].$$
The apparent "stiffnesses" $\stiffnessone$ and $ \stiffnesstwo$ are a measure of the relative stiffness of each lamellae. Thus, they are proportional to the number of fibers in the lamellae direction and hence to the coefficients $C_1$ and $C_2$ (Eq.~\eqref{I_minimised}). Thus, there is a proportionality factor $\stiffnessfibrilapparent$ such that:
\begin{equation}
k_i = \stiffnessfibrilapparent C_i  \\
\label{stiffnesses_relationship}
\end{equation}
Finally, we can define an effective "stiffness" $k_{lam} = l_u  \stiffnessfibrilapparent$, so that the energy function becomes:
\begin{equation}
	\displaystyle  \dpsiAnisoi \thetaphi := \frac{1}{2} C_i k_{lam} ( \frac{ \elongationfiber_i}{\lambda_{u,i}  } - 1) _{+}^2, \quad \forall i \in [1:2].
	\label{dpsianiso_all_reduced}
\end{equation}
and so it leaves only a global constant parameter $k_{lam}$. \\
The last parameters are the unfolding stretches $ \unfoldingelongationfibone, \unfoldingelongationfibtwo$. The "unfolding" elongations are supposed to depend on the dispersion of the lamellae. Indeed, the more the lamellae are stretched in the reference configuration (i.e. the closer the "unfolding" elongation is to 0), the more the lamellae are aligned, therefore the less they are dispersed (i.e. the greater the $\inplanefiberdensity$). On the contrary, the less the lamellae are stretched in the reference configuration (i.e. the closer the reference length is to the "unfolding" length), the less the lamellae are aligned, therefore the more they are dispersed (i.e. the smaller the $\inplanefiberdensity$). In a first approach, they are considered to be linearly inversely proportional $\unfoldingelongation = a / \inplanefiberdensity + b$, with coefficients $a$ and $b$ to be determined thanks to the limits:
\begin{equation}
\displaystyle \unfoldingelongationmin = \frac{a}{\inplanefiberdensitymax} + b, \quad \text{and} \quad \displaystyle \unfoldingelongationmax = \frac{a}{\inplanefiberdensitymin} + b
\label{unfolding_elongation_max_min}
\end{equation}
which makes for two news independent parameters $\unfoldingelongationmax$ and $\unfoldingelongationmin$ the maximum and minimum unfolding elongation of the lamellae in the whole cornea, to be determined experimentally.

Anisotropic contribution (Eq.~\eqref{psi_aniso_continuous}) finally reduces to
\begin{equation}
	\displaystyle \psiAniso \vertjoliwriting{ = \int_{\theta = 0}^{\pi} \int_{\phi = 0}^{2 \pi}  \sum_{i=1}^{2} \frac{1}{2} C_i k_{lam} \big( \frac{ \elongationfiber_i \thetaphi }{ \lambda_{u,i}  } - 1 \big)_{+}^2   \frac{ e^{ \kappa_{ip,i} \cos(2 (\phi-\mu_i) ) } e^{ \kappa_{t,i} \cos(2\theta) } } { C^{lam}_i }  \sin \theta \text{d} \theta  \text{d} \phi    }
\label{psi_aniso_reduced}
\end{equation}
with only three unknown global parameters left $\unfoldingelongationmax$, $\unfoldingelongationmin$ and $k_{lam}$.

Table~\ref{independant_parameters} summaries the independent global parameters used in the model, the constitutive equations where they appear and the values determined to reproduce the experimental data from \cite{elsheikh_biomechanical_2008} and \cite{mcmonnies_corneal_2010}.

\begin{tableth} 
	\begin{tabular}{|c|c|p{5.5cm}|p{2.3cm}|c|}
		\hline
		Parameter notation & Energy function &  Parameter description & Equation & Value \\
		\hline
		$\bleufoncecwriting{\apparentpsiisoIoneparameter} $  & \multirow{5}{*}{} &   Matrix stiffness  & Eq. \eqref{psi_iso}, \eqref{kappa_1_apparent} & $60$Pa  \\
		\cline{1-1} \cline{3-5} 
		$ \alpha $  & $\psiIso$ &   Proportional factor between the two matrix parameter & Eq. \eqref{Iiso2} & $1/4$ \\
		\hline
		$\rosewriting{K}$ & $\psiVol$ &  Hyperelastic bulk & Eq. \eqref{psi_vol} & $80$ kPa\\
		\hline
		$k_{lam}$ & \multirow{5}{*}{} & Apparent stiffness of a collagen lamellae for a given length & Eq. \eqref{dpsianiso_all_reduced}, \eqref{psi_aniso_reduced} &  $65$ Pa\\
		\cline{1-1} \cline{3-5} $\unfoldingelongationmax$ &  $\psiAniso$ &   Maximum "unfolding" elongation $\unfoldingelongation$ in the reference configuration & Eq. \ref{unfolding_elongation_max_min}, \eqref{psi_aniso_reduced} & 1.0245 \\
		\cline{1-1}  \cline{3-5} $\unfoldingelongationmin$  & & Minimum "unfolding" elongation $\unfoldingelongation$ in the reference configuration & Eq. \eqref{unfolding_elongation_max_min}, \eqref{psi_aniso_reduced}  & 1.0195\\
		\hline
	\end{tabular}
	\caption[Summary of the global parameters of the model]{Summary of the global parameters of the model, their contribution, where they appear, and their values determined by simulating an inflation test to reproduce the data from \cite{elsheikh_biomechanical_2008}. \label{independant_parameters}}
\end{tableth}

Once we have simplified the model by reducing the number of independent parameters, we use a finite element code - MoReFEM - developed at Inria by the M$\Xi$DISIM team \cite{gilles_gitlab_nodate} to solve Eq.~\eqref{final_weak_formulation_Sigma}. The Galerkin method is used to do the spatial discretization, using Q1 hexaedric finite elements. To compute the anisotropic part of the 2$^{nd}$ Piola-Kirchhoff tensor at Gauss points, a numerical quadrature is used for the integral (Eq.~\eqref{psi_aniso_continuous}) on the microsphere using a uniform rule with 20 equally distributed points for the in-plane angle $\phi$ and the Gauss-Hermite quadrature rule with $5^{th}$ order polynomial and 5 quadrature points for the out-of-plane angle $\theta$. Two loading conditions are used:
\begin{itemize}
	\item Loading from $2$ mmHg to $160$ mmHg to mimic the \textit{ex-vivo} experiment of Elsheik et al. \cite{elsheikh_biomechanical_2008} on human cornea under pressure: we use this to calibrate the model.
	\item Loading from $15$ mmHg to $30$ mmHg to mimic the \textit{in-vivo} experiment of McMonnies and Boneham \cite{mcmonnies_corneal_2010}: we use this to investigate the origin of the keratoconus.
\end{itemize}

\subsection{Stress-free configuration \label{Stres_free} }

To numerically solve Eq.~\eqref{final_weak_formulation_Sigma}, we need to start from a stress-free configuration. However, the patient-specific geometry is obtained under physiological intra-ocular pressure (IOP). As IOP was not determined during this clinical acquisition, we assume that it is the mean IOP of healthy individuals ($14.5$ mmHg \cite{hashemi_distribution_2005}). We then use the patient-specific configuration $\Omega_{physio}$ (associated to the positions $\vect{x}_{physio}$) as the target of a shooting method to determine the stress-free configuration. Starting from an assumed reference configuration $(\Omega_0 , \vect{\xi})$, the procedure is the following:\\
\begin{algorithm}
	\caption{Computation of the stress-free configuration}
	\begin{algorithmic}
\STATE	\textbf{Step 1}	 - Computation of the deformed configuration under IOP pressure $(\Omega_p, \vect{x}_p)$\\
\STATE	\textbf{Step 2}	 - Determine the differences $\displaystyle \vect{\Delta}_{\vect{x}} = \vect{x}_p-\vect{x}_{physio}$.\\
\STATE	\textbf{Step 3}	 - While any of the differences $|\vect{\Delta}_{\vect{x}} |$ is larger than a tolerance (taken at $10^{-6}$mm), update the reference configuration by $\vect{\xi}_{new} = \vect{\xi} - \displVect $. Otherwise, we consider that we have found the reference configuration.
	\end{algorithmic}
\end{algorithm}

Figure~\ref{corrected_mesh}c presents the two meshes used in the algorithm for a stage 4 keratoconic cornea. The pink one is the corrected mesh under physiological pressure $\Omega_{physio}$ and the blue one corresponds to the associated stress-free configuration $\Omega_{0, stress-free}$ mesh (for $P=0$ mmHg): the two being barely distinguishable. Note that the reference configuration needs to be updated each time you change any mechanical parameter of the model.

\subsection{simK determination}

To compare our data with McMonnies and Boneham \cite{mcmonnies_corneal_2010}, we computed the simK of our cornea at different pressures. The simK is the diopter (D) associated to the steepest meridian of the cornea as identified at a small radius ($r = 1$ mm - see Fig.~\ref{simK_computation}). To compute the simK, we fit the biconic equation (Eq.~\eqref{biconic_equation}) on the deformed anterior surface inside a $1$mm radius from the apex. We obtain the two radii for each level of pressure and from them we can compute the diopter D using the steepest one:
\begin{equation}
	D(P)= simK(P) = \frac{ n_{aqh} - n_{air} }{R_{steep(P)}}
\end{equation}
where $R_{steep}$ is the radius of the steepest meridian and $n_{aqh}$ and $n_{air}$ are the refraction indexes of the aqueous humor and air (taken at 1.3375 and 1.0000 respectively).

 \begin{figureth}
 	\includegraphics[width = 0.7\linewidth]{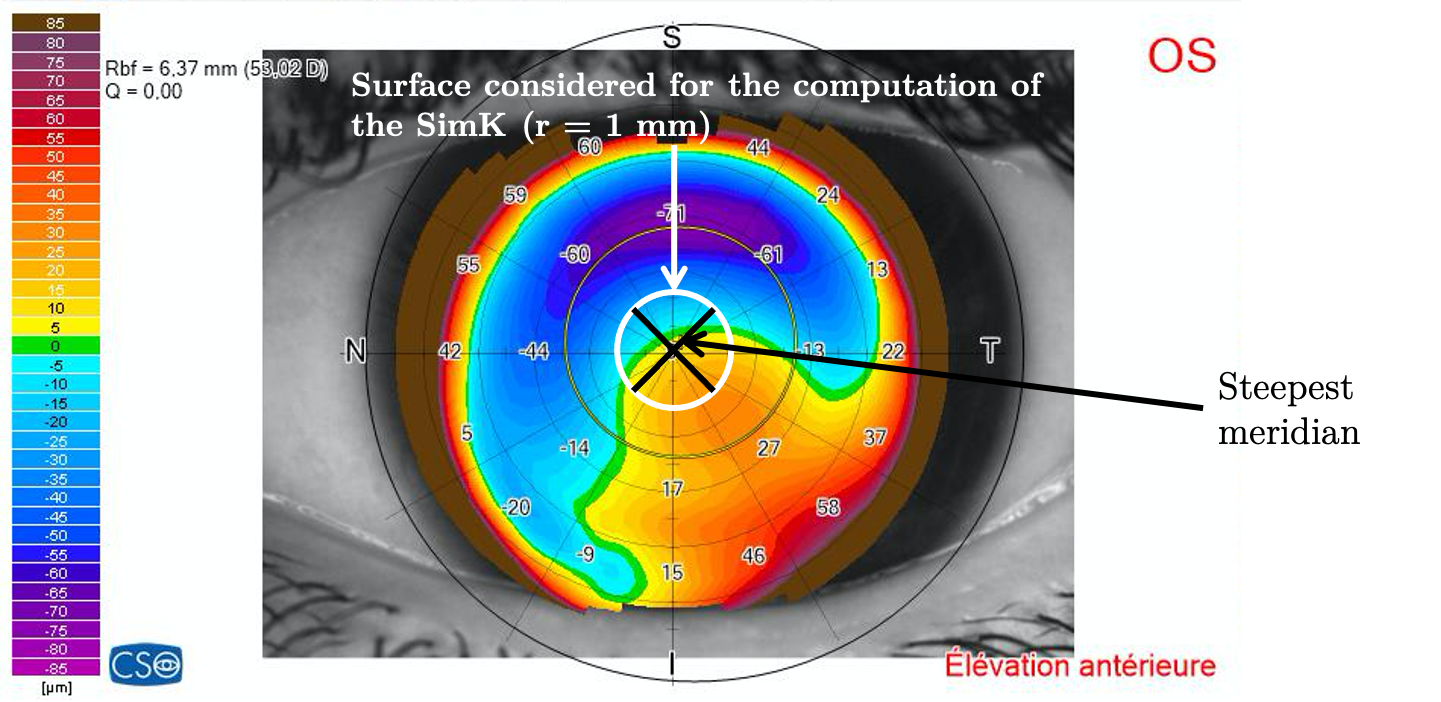}
 	\caption[SimK Computation]{\label{simK_computation} Example of the considered surface used to compute the SimK. A subregion of $1$ mm in radius of the anterior surface is fitted by a biconic function. The steepest meridian is used to compute the SimK.}
 \end{figureth}
	\section{Results}

We first determine the values of our model parameters by reproducing experimental data on \textit{ex-vivo} inflation assays \cite{elsheikh_biomechanical_2008}: these parameters will be our "reference" parameters used to investigate the origin of the keratoconus.

\subsection{Parameter estimation \label{parameter_estimation}}
	We simulated the experiment by Elsheik et al. \cite{elsheikh_biomechanical_2008}. To do so, we used the stress-free geometry $\Omega_{0, stress-free}^{ref}$ of a healthy cornea and applied a pressure from $0$ to $160$ mmHg while determining the apex displacement. Figure~\ref{apical_displacement_plus_minus_1_percent} shows the envelope of the experimental data (in pink), which comes from inter-cornea variability. The triangular markers are our simulation using the "reference" parameters (see Table \ref{independant_parameters}), obtained after manual calibration. 
	
	We have then varied each parameter independently by $1\%$. The most sensitive parameters are the unfolding stretches $\unfoldingelongationmin$ and $\unfoldingelongationmax$ (the results for the other parameters are presented in appendix~\ref{sensitivity_analysis}, Fig.~\ref{apical_displacement_all}). Figure~\ref{apical_displacement_plus_minus_1_percent} shows that an increase (resp. decrease) of both the unfolding stretches by $1\%$ moves the pressure vs apex displacement curve to the right (resp. to the left), well outside the experimental data range. Unfolding stretch corresponds to the stretch above which the lamellae start to respond elastically. As $\lambda_u > 1$, the lamellae in the reference configuration are folded and do not contribute to the tissue rigidity. Once they become activated, the tissue becomes much stiffer. This explains why a change in the unfolding stretch leads to a shift of the pressure vs apical displacement curve: increasing the unfolding stretch will elongate the heel region, without changing the linear part so much. 
	
\begin{figureth}
	\includegraphics[width = 1.1\linewidth]{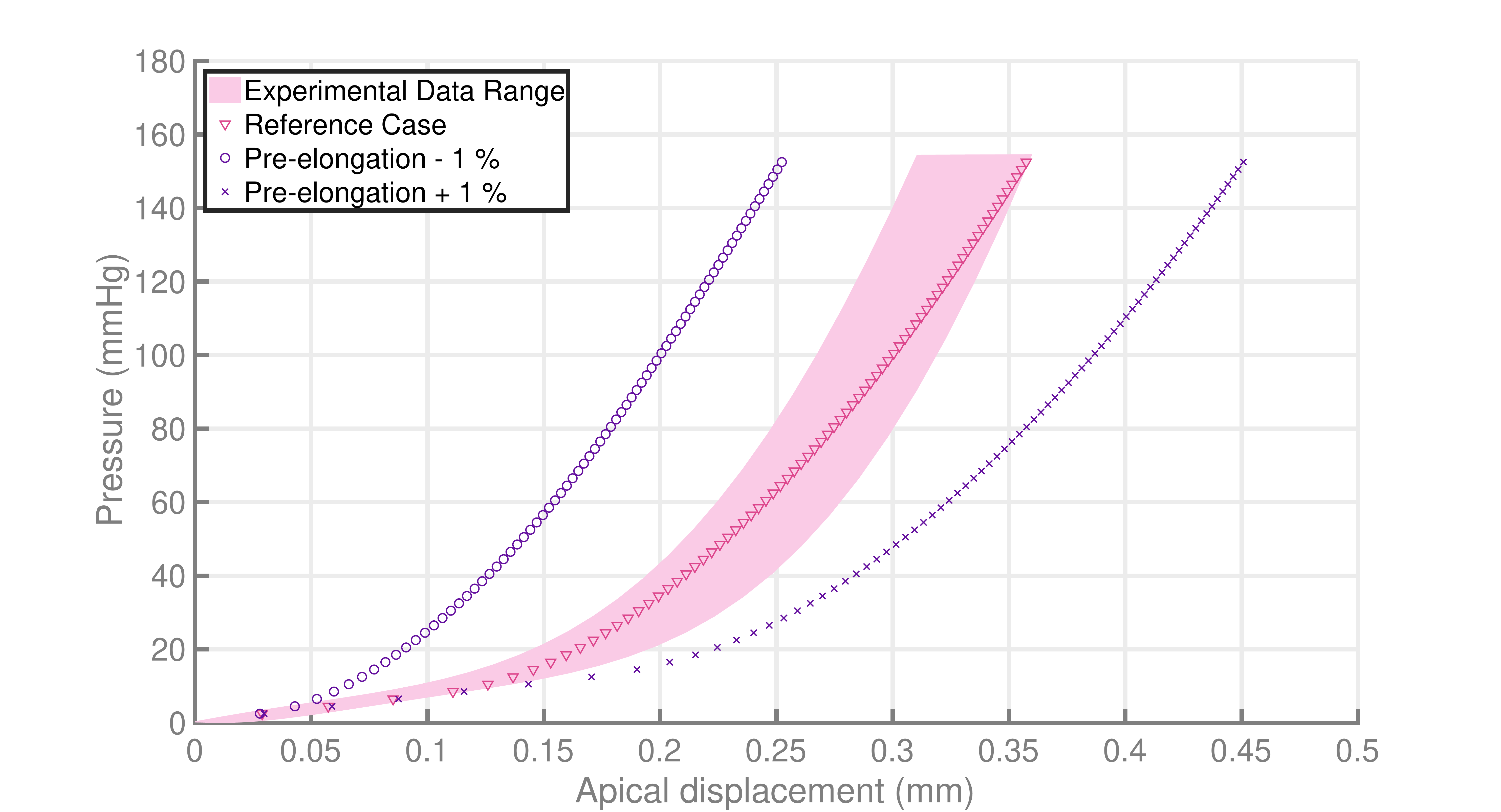}
	\caption[Pressure with apical displacement for different cases of limits values of $\protect \unfoldingelongation$]{\label{apical_displacement_plus_minus_1_percent} Pressure with apical displacement for three different $\protect \unfoldingelongation$. Pink zones: envelopes of the experimental data from \cite{elsheikh_biomechanical_2008}.'$\nabla$': reference case. 'o': $1\%$ decrease of the $\unfoldingelongation$. 'x': $1\%$ increase of the $\unfoldingelongation$. }
\end{figureth}

\subsection{Keratoconus: geometrical and mechanical effect \label{keratoconus_geometry}}

To distinguish between mechanical and geometrical origin of keratoconus, we first simulated a healthy and a stage 4 keratoconic cornea with "reference" mechanical parameters, and compared with the observations from McMonnies and Boneham \cite{mcmonnies_corneal_2010}. They showed that the simK of the healthy corneas does not change significantly for a change of intra-ocular pressure in the range of $15-30mmHg$ whereas the simK of keratoconic corneas increases of 2 diopters. Figure~\ref{simK} shows the simulated keratometry (or simK) as a function of the applied pressure: for the "reference" parameters ($\nabla$ symbols, see table~\ref{studies_parameters}) in both healthy (pink) and keratoconic (purple) corneas, the simK does not change significantly (less than $0.5$ diopter). This implies that a modification of the mechanical properties is needed to reproduce the keratoconus response.

Then, we modified the mechanical parameters to obtain a change of keratometry of 2 diopters, by a manual adjustment. We modified separately the non-fibrillar matrix stiffness ($\apparentpsiisoIoneparameter$), the distributed fibril stiffness ($C_i*k_{lam}$), or the pre-elongation ($\lambda_u$). The only parameter that gives a significant change of diopter without changing of order of magnitude is the fibril stiffness $k_{lam}$, the mean values of distributed lamellae stiffnesses $(C_1*k_{lam})$ and $(C_2*k_{lam})$ decreasing by around 40 and 30\% respectively. Table~\ref{studies_parameters} gives the changed parameters of each simulation.
To obtain a change of $1$ diopter by weakening the matrix, a two orders of magnitude change was needed on $\apparentpsiisoIoneparameter$, and no set of parameters was found to have a change greater than $0.3$ diopter thanks to a variation of the pre-elongation parameters $\lambda_u$. Figure~\ref{simK} shows the simK variation with pressure of the reference and weakened fibril stiffness cases. Our results show that the keratoconus pressure response can easily be captured by a change in the mechanical behavior, even if we changed the parameters slightly differently for healthy and keratoconic corneas.

\begin{figureth}
	\includegraphics[width = 0.9\linewidth]{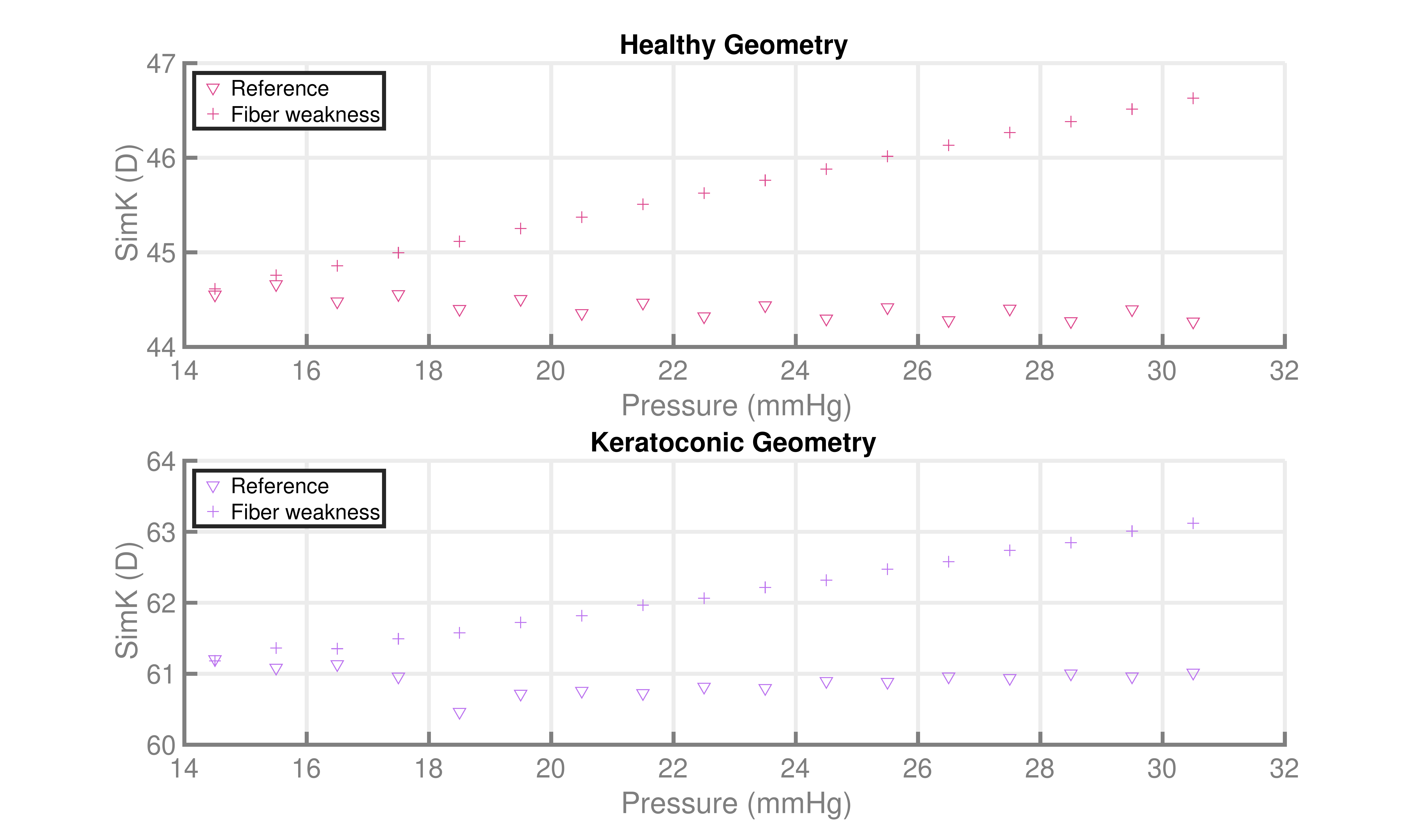}
	\caption[SimK Computation]{\label{simK} Computation of the SimK for the reference and fiber weakness cases considered in table~\ref{studies_parameters} with healthy (up) and keratoconic (down) geometries. Modifying the value of the mechanical parameters of the anisotropic part of the cornea, a variation of 2 diopters can be observed.}
\end{figureth}

Figure~\ref{CS_diopter_all_15mmHg} presents the stresses in the Nasal-Temporal (NT) and Superior-Inferior (SI) directions for healthy and keratoconic geometries, without and with mechanical weaknesses at physiological pressure. The pattern at the boundary is due to the highly rigid boundary condition, and is heterogeneous in the thickness. Both healthy and keratoconic corneas show a higher concentration of the stress in the central region of the anterior surface (even higher in the keratoconic case),  whereas the stress in the posterior surface is quite homogeneous. This means that the geometry has a strong impact on the stress, even if it does not affect the keratometry response. On the contrary, modifications of the mechanical parameters do not affect the pattern strongly - mainly smoothing it. This indicates that the stress distribution is mostly due to the fiber distribution, except at the vicinity of the corneal boundary.

\begin{figureth}
	\includegraphics[width = 1.1\linewidth]{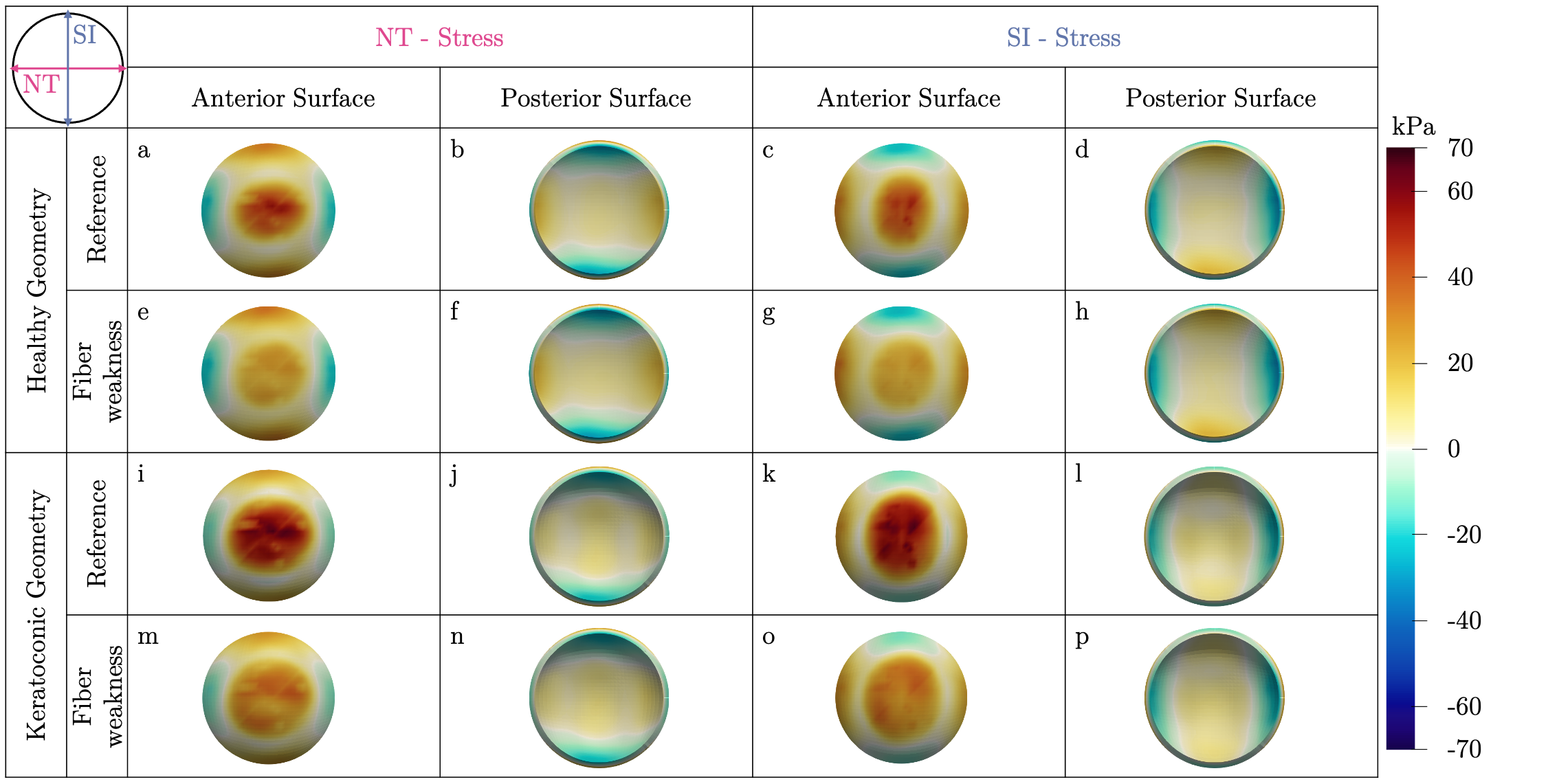}
	\caption[Cauchy stress at physiological pressure for different cases of mechanical weaknesses in the cornea]{\label{CS_diopter_all_15mmHg} Cauchy stress at physiological pressure for different cases of mechanical weaknesses in the cornea with healthy and keratoconic geometries. Fig.~\ref{CS_diopter_all_15mmHg}a-h: Naso-Temporal and Superior-Inferior stresses for the healthy geometry (a-d: reference case and e-h: case of the fibril weakness with an increase of 2 diopters between 15 and 31 mmHg) on the anterior and posterior surfaces. Fig.~\ref{CS_diopter_all_15mmHg}i-p:  Naso-Temporal and Superior-Inferior stresses for the keratoconic geometry on the anterior and posterior surfaces (i-l: reference case and m-p: case of the fibril weakness with an increase of 2 diopters between 15 and 31 mmHg). }
\end{figureth}

\subsection{Induced keratoconus \label{induced_keratoconus}}

So far, we have separated the problem of the geometry and of the mechanical parameters: we have chosen either the healthy parameters and changed the geometry, or chosen an observed geometry and modified the mechanical parameters. In both cases, we show that the change in diopter associated with keratoconus response cannot be explained by the change in geometry but can be reproduced by a decrease in the mechanical properties, in particular of the fiber rigidity. To do so, we started from an observed geometry, and simulated a stress-free configuration, obtained such that it reproduces at physiological pressure the observed geometry, for the chosen set of mechanical parameter. This means that the keratoconic cornea has a stress-free configuration which is different from the healthy cornea. Here, we ask ourselves what will be the geometry of a cornea under pressure if we use on the healthy-stress cornea the keratoconic mechanical parameters: we would like to see if the change of mechanical parameters is able to recreate the keratoconic geometry. 

We first determine the stress-free configuration of our reference case (healthy geometry, with reference mechanical parameters), and simulated the response of the cornea at different pressures for weakened fibril stiffness corresponding to a $2$ diopter increase. 

Figure~\ref{SimK_RefMesh20} shows the computed SimK for this new case. We also reproduced the simulation of the reference case, which leads to a constant SimK (see Fig.~\ref{simK}). The decreased mechanical properties lead to a higher SimK at physiological pressure than for the reference case, although it is smaller than the one for the simulation starting from keratoconic stress-free configuration (around $61$ D). This reflects the fact that a different stress-free configuration will lead to a different geometry under pressure, and is in line with stage-1 keratoconus based on Krumeich's classification \cite{naderan_histopathologic_2017}. We also observe an increase of $2$ diopters, consistent with a keratoconic response.

\begin{figureth}
	\includegraphics[width = 0.9\linewidth]{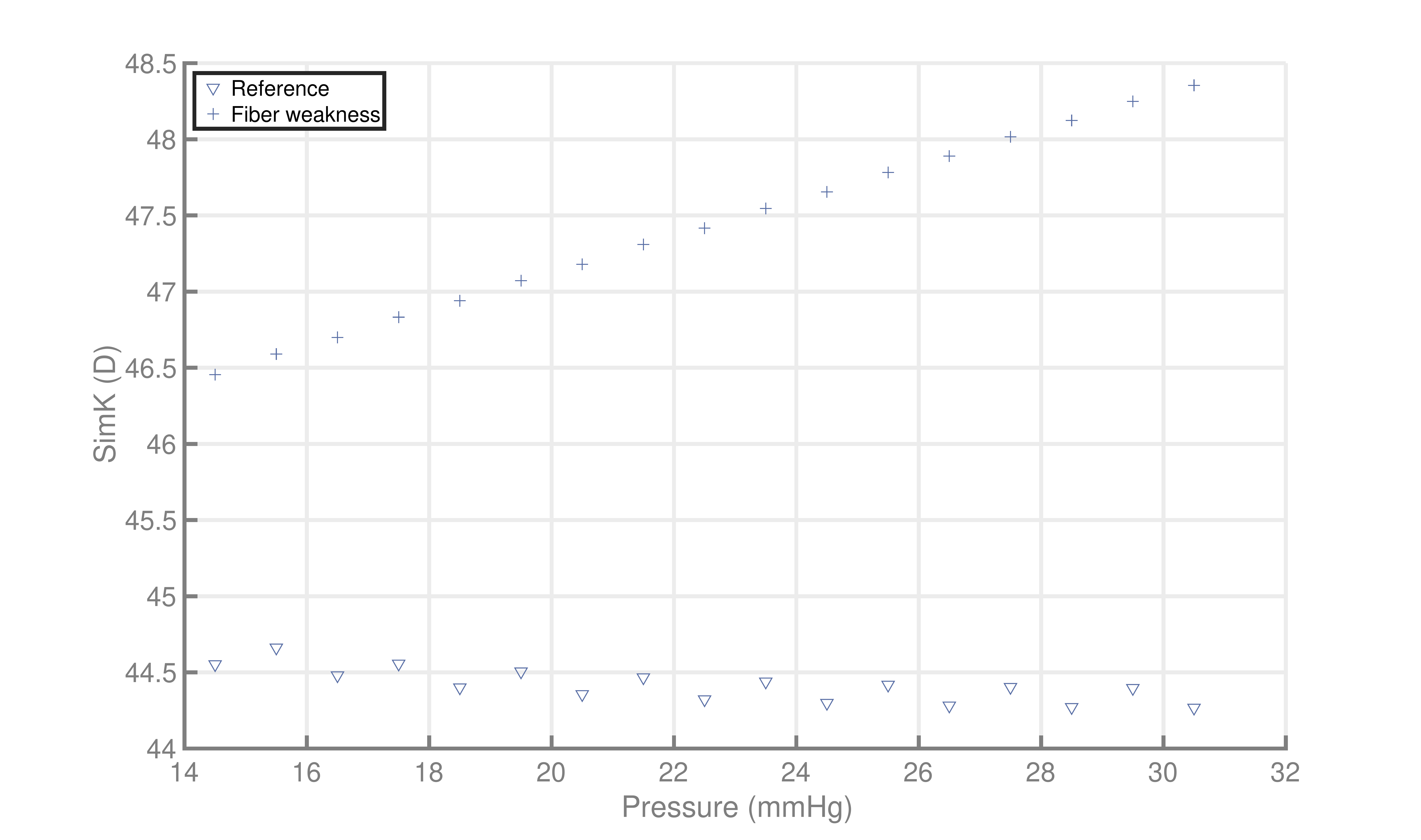}
	\caption[Simulated keratoconus]{\label{SimK_RefMesh20} SimK computed for the different cases of mechanical weakness on the reference stress-free configuration.}
\end{figureth}

Figure~\ref{CS_diopter_healthy_mesh20_15mmHg} presents the NT ans SI stress distributions for the reference and mechanical weakness cases. The distributions of stresses are very similar to those in Fig.~\ref{CS_diopter_all_15mmHg}, in agreement with our previous observation that this stress pattern is more controlled by the fiber distribution than by the cornea geometry.

\begin{figureth}
	\includegraphics[width = 1.0\linewidth]{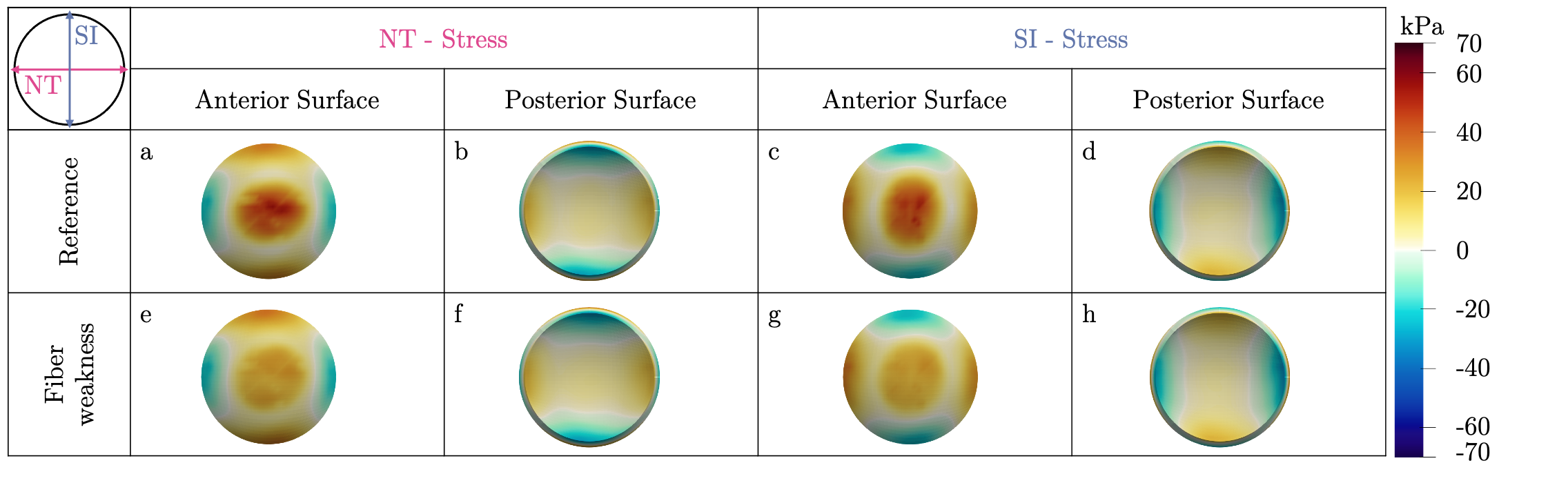}
	\caption[Cauchy stress at physiological pressure for different cases of mechanical weaknesses in the cornea]{\label{CS_diopter_healthy_mesh20_15mmHg} Stress at physiological pressure with reference parameters and mechanical weakening of the cornea with healthy geometry and stress-free configuration of the reference case used for every computation.
		Fig.~\ref{CS_diopter_healthy_mesh20_15mmHg}a-d: Naso-Temporal and Superior-Inferior stresses for the reference case a, Fig.~\ref{CS_diopter_healthy_mesh20_15mmHg}e-h:  Naso-Temporal and Superior-Inferior stresses for case of the fibril weakness with an increase of 2 diopters between 15 and 31 mmHg.}
\end{figureth}

Figure~\ref{map_elevation_RefMesh20_all} show the elevation maps obtained at physiological pressure and at $P = 30$ mmHg, for the reference case and for the weakened mechanical properties. The fibril weakening does not lead to a major change of the elevation, but we can see that in the posterior surface, the elevation in the central region is higher than in the reference case (it is even clearer at 30 mmHg), which can lead to the suspicion of a very early stage of a keratoconus. Those results are coherent with the value of the SimK at physiological pressure previously computed and tend to indicate that the keratoconus may appear following a weakening of the anisotropic part of the cornea. On the other hand, elevation maps do not show an off-centered elevation (neither an off-centered thinning on thickness maps is seen) that could lead to suspect a keratoconus \cite{duncan_assessing_2016, belin_scheimpflug_2013}. Indeed the quasi-incompressibility of the cornea does not allow for a significant change in the cornea geometry with a thinning of the cone region, thus it cannot change to become an advanced stage keratoconic cornea,  although the change of diopter - and thus the change of curvature radii - is coherent with a keratoconus.

\begin{figureth}
	\includegraphics[width = 1.0\linewidth]{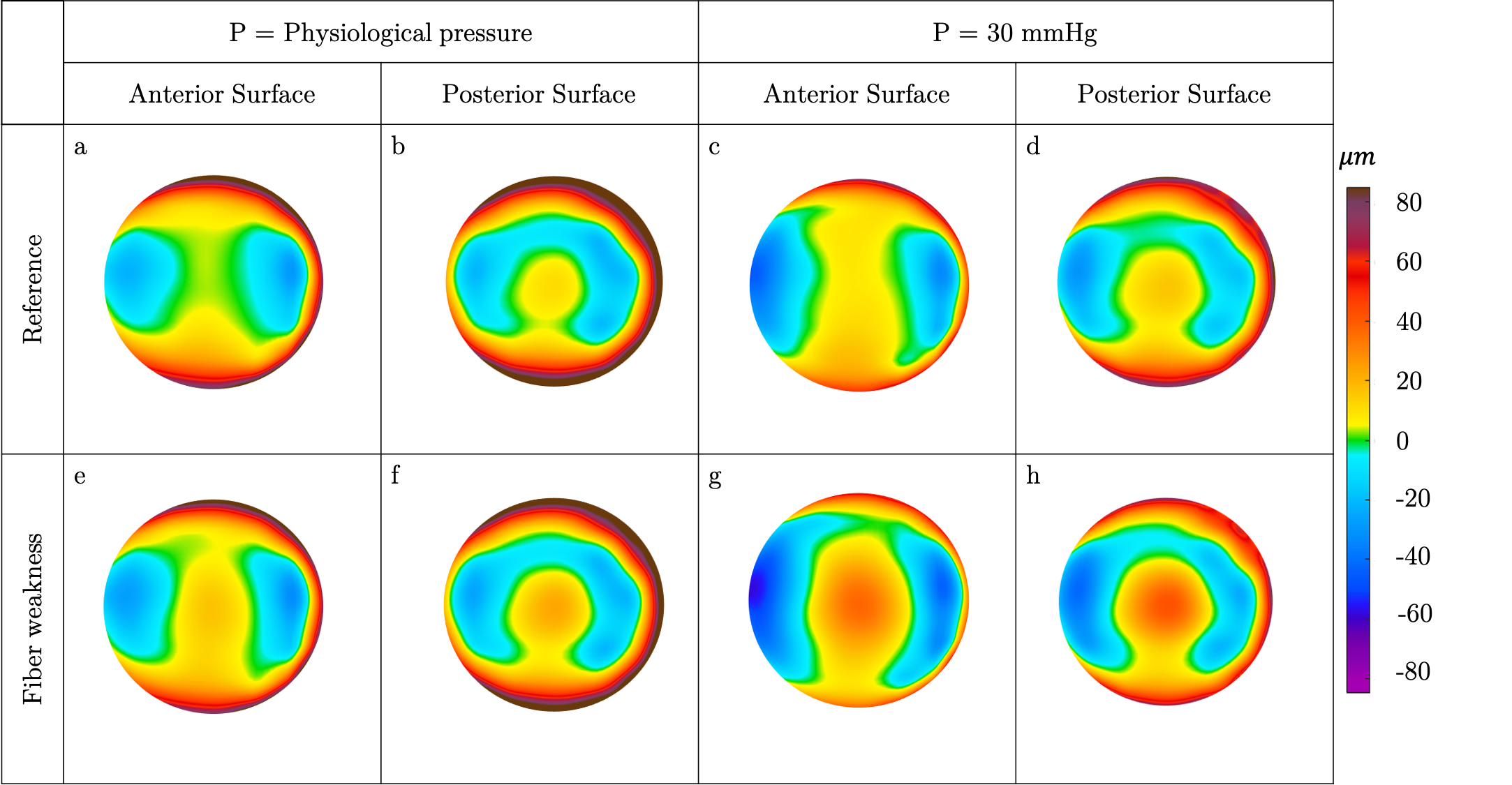}
	\caption[Elevation maps]{\label{map_elevation_RefMesh20_all} Anterior and posterior elevation maps with respect to the best fit spheres for the reference (a-d) and weakened fiber (e-f) cases at physiological pressure (a,b,e,f) and for $P=30$ mmHg (c,d,g,h).}
\end{figureth}

	\section{Discussion}

To investigate the origin of keratoconus, we have compared the effects of a change in geometry and of a change in mechanical properties. To do so, we constructed a patient-specific mesh, which reproduces the geometry measured in clinic. We have built a multi-scale model, which contains explicitly the different contributions (fibrils, isotropic matrix, etc.), but it was not possible to obtain patient-specific data for these parameters. The collagen organization was obtained from experimental observations (X-ray \cite{aghamohammadzadeh_x-ray_2004} or SHG \cite{winkler_three-dimensional_2013}). The different mechanical stiffnesses were manually calibrated to reproduce the reported data \cite{elsheikh_biomechanical_2008}. As we have access only to the displacement of the apex in human cornea, with a variability between corneas, we did not try a proper identification. This implies that our "reference" set of parameters may not be unique. Corneal strain maps have been measured on other animals (as bovine \cite{boyce_full-field_2008}), but then the keratoconus geometry is not available on the same animal.

We have tested the influence of small variations of each mechanical parameter, and we observed that the most sensitive one is the unfolding stretch, i.e. the stretch at which the fibrils start to generate force. Associated with our observation that the stress distribution corresponds to the fibril distribution (Fig.~\ref{CS_diopter_all_15mmHg}), this supports the idea that the forces in the cornea are mainly due to the fibrils, and only partly to the isotropic matrix or the volume variation. Note that the fibers become more and more unfolded as the pressure increases above physiological pressure, contributing to the increase of the tissue stiffness (see Fig.~\ref{apical_displacement_plus_minus_1_percent}). SHG observations of the lamellae show straight fibrils \cite{latour_vivo_2012, benoit_simultaneous_2016}. It may be explained by the fibril organization at smaller scale \cite{bell_hierarchical_2018}. In any case, it implies that the fibril tensions play a major role in the corneal response, which could have an impact on the recovery of the cornea after a laser surgery.

We simulated the inflation of a cornea with a keratoconic geometry. Using directly our reference mechanical parameter fails to reproduce the reported variation of keratometry during the inflation test \cite{mcmonnies_corneal_2010}. This shows that keratoconic geometry alone (thinner cornea) is not enough to have a keratoconic behavior. However, a 30 to 40 \% decrease in the average fiber stiffnesses allows our model to reproduce the $2$ diopters variations, even for healthy geometries. Thus, our approach shows that mechanical weakening, contrary to the geometry, is able to reproduce the keratoconus changes in SimK, emphasizing the importance of mechanical weakening on the keratoconic response. The weakened parameters reproducing the keratoconus behavior (see Table~\ref{studies_parameters}) indicate that it requires a relatively small decrease of the fibril stiffness to obtain a $2$ diopter variation. This points toward the key role of the collagen lamellae in the development of the keratoconus, in agreement with the proposed treatments by the addition of cross-links.

By using the weakened mechanical parameters on the healthy stress-free configuration (see Fig.~\ref{map_elevation_RefMesh20_all}), we were able to reproduce partly a keratoconic shape at physiological pressure. This again supports the idea that the primary motor of the keratoconus is a weakening of the collagen fibrils, consistent with the disorganization of the lamellae observed in \cite{meek_changes_2005}. However, the obtained shape is not the one of a real keratoconus, with a large elevation peak slightly off-centered. This may come from our quasi-incompressibility assumption, which prevents a thinning of the cornea. But more likely, to go further in the modeling of the keratoconus, we need a better understanding of the remodeling going on inside the tissue.
	\section{Conclusions}

In this paper, we have built a multi-scale model of the cornea, coupled to a patient-specific geometry to investigate the origin of the keratoconus. We have first used our model to reproduce the pressure versus apex displacement curve from Elsheikh et al. \cite{elsheikh_biomechanical_2008} and determined a reference set of mechanical parameters, describing a healthy cornea. We show that the central element of the mechanical response is the one of the fibrils, and in particular their prestretch. 

Our simulation of cornea with keratoconic geometry but healthy mechanical parameters shows that the geometry change is not able to reproduce the response of keratoconic cornea to an increase of the intraocular pressure \cite{mcmonnies_corneal_2010}. In fact, we showed that the keratoconic response is well reproduced when the mechanical properties are altered, whatever the initial geometry, and that the main component involved in this response is the lamellae stiffness. The lamellae weakening is even sufficient to obtain a shape resembling an early-stage keratoconus.

Although they could be completed by a better description of the induced remodeling, our simulations show the importance of a fine measurement of the mechanical properties in the understanding and diagnosis of keratoconus.

\paragraph*{Acknowledgment}
We kindly thank A. Pandolfi for providing the 3D mesh code, K. M. Meek and S. Hayes for the corneal X-ray experimental data and J. Knoeri and V. Borderie for providing elevation and thickness maps.

\paragraph*{Authors' contribution} C.G. and P.L.T designed the model. C.G. and J.D. implemented the model. C.G. performed the simulations, and analyzed the results. C.G. and J.M.A. discussed the results. J.M.A. supervised the research. All authors read and approved the manuscript.

\paragraph*{Competing interests}
The authors declare no competing interests.

\paragraph*{Financial disclosure}
No funding has been received for this article.

	\newpage
	\bibliographystyle{plain}
	\bibliography{RefPaperModel.bib}
	\begin{appendix}
		\pagenumbering{Roman}
		\section{Sensitivity analysis \label{sensitivity_analysis}}

\begin{figureth}
	\includegraphics[width = 1.1\linewidth]{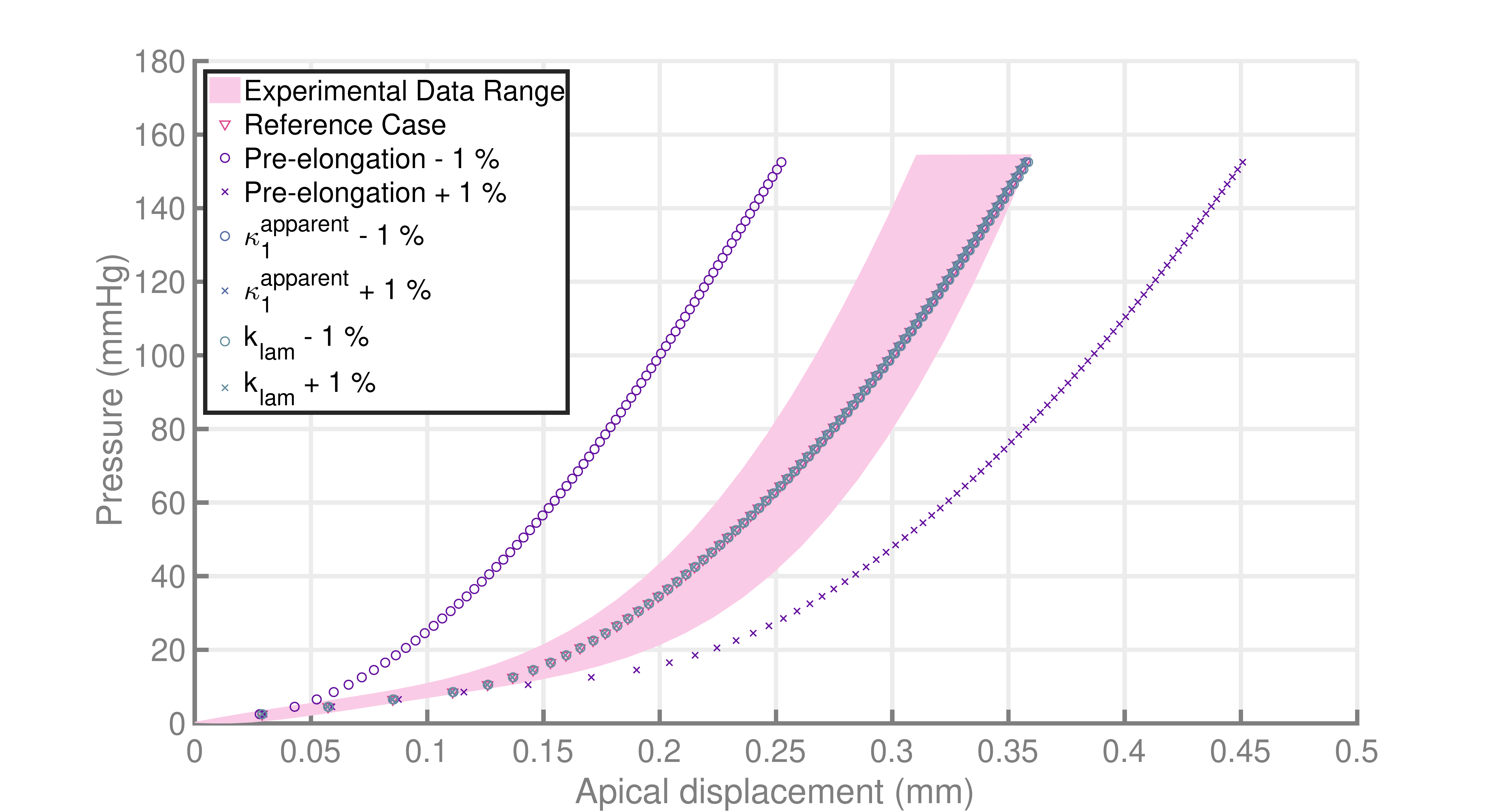}
	\caption[Pressure with apical displacement for sensitivity analysis cases ]{\label{apical_displacement_all} Pressure with apical displacement for sensitivity analysis cases. Pink zones: envelopes of the experimental data from \cite{elsheikh_biomechanical_2008}.'$\nabla$' markers curve: reference case. 'o' violet markers curve: $1\%$ decrease of the $\unfoldingelongation$. 'x' violet markers curve: $1\%$ increase of the $\unfoldingelongation$. 'o' blue markers curve: $1\%$ decrease of the $\kappa_{1}^{apparent}$. 'x' blue markers curve: $1\%$ increase of the $\kappa_{1}^{apparent}$. 'o' green markers curve: $1\%$ decrease of the $k_{lam}$. 'x' green markers curve: $1\%$ increase of the $k_{lam}$.}
\end{figureth}

\section{Mechanical parameters used in the computation}

\renewcommand{\arraystretch}{1.5}
\begin{tableth}
	\begin{tabular}{|p{5cm}|p{4.6cm}|p{4.6cm}|}
		\hline
		Geometry & Healthy (associated stress-free configuration) & Keratoconic (associated stress-free configuration )  \\
		\hline
		Ref = no mechanical weakness &  RefH ($\displaystyle \Omega_{0}^{RefH}$ ) & RefK ($\displaystyle \Omega_{0}^{RefK}$ )  \\
		\hline
		ElongM1 = Pre-elongation minus one percent  & ElongM1 ($\displaystyle \Omega_{0}^{ElongM1}$ ) & /  \\
		\hline
		ElongP1 = Pre-elongation plus one percent  &  ElongP1 ($\displaystyle \Omega_{0}^{ElongP1}$ ) & /\\
		\hline
		Fib2 = Mechanical weakness on the lamellae leading to a 2 diopters change   &  Fib2H ($\displaystyle \Omega_{0}^{Fib2H}$) and Fib2H2 ($\displaystyle \rosewriting{\Omega_{0}^{RefH}}$ ) & Fib2K ($\displaystyle \Omega_{0}^{Fib2K}$ ) \\
		\hline
	\end{tabular}
	\caption[Keratoconic cases]{Cases considered in the mechanical study of keratoconic cornea (Sec.~\ref{keratoconus_geometry} and~\ref{induced_keratoconus}). The reference case for healthy geometry (RefH) corresponds to the ones calibrated on Elsheik's group data (see Sec.~\ref{parameter_estimation}). Between brackets are noted the stress-free meshes $\displaystyle \Omega_{0, stress-free}$ used for each computational cases. \label{all_cases_names} }
\end{tableth}

\renewcommand{\arraystretch}{1.5}
\begin{tableth}
	\begin{tabular}{|p{5.78cm}|c|c|c|c|c|}
		\hline
		\diagbox{Parameter}{Case considered} & RefH / RefK & ElongM1 & ElongP1  & Fib2H / Fib2H2 &  Fib2K \\
		\hline
		Average of distributed isotropic coefficient $\displaystyle\bleufoncecwriting{\kappa}_1 (MPa)$ &  0,97 &  0,97  &  0,97  &  0,97  &  0,97 \\
		\hline
		Minimum "unfolding" elongation $\displaystyle \lambda_{u,min}$  & 1,0195 & \rosewriting{1,0093}  &\rosewriting{1,0297 }   & 1,0195 & 1,0195 \\
		\hline
		Maximum "unfolding" elongation $\displaystyle \lambda_{u,max}$  & 1,0245 & \rosewriting{1,0143} & \rosewriting{1,0347} & 1,0245  & 1,0245 \\
		\hline
		Average of distributed anisotropic coefficient $\displaystyle \vertjoliwriting{C_1*k_{lam}}$ (MPa) & 7,15 & 7,15  & 7,15 & \rosewriting{4,10} & \rosewriting{4,11}\\
		\hline
		Average of distributed anisotropic coefficient $\displaystyle \vertjoliwriting{C_2*k_{lam}}$ (MPa) & 18,70 & 18,70  & 18,70 & \rosewriting{12,43}  & \rosewriting{12,76} \\
		\hline
	\end{tabular}
	\caption[Reference parameters]{Mechanical parameters used in the different computations on the cornea. The different cases are presented in Table~\ref{all_cases_names}. For the distributed parameters ($\displaystyle\bleufoncecwriting{\kappa_1}$, $\vertjoliwriting{C_1*k_{lam}}$ and $\vertjoliwriting{C_2*k_{lam}}$) the average values on all over the cornea are given.  \label{studies_parameters} }
\end{tableth}

	\end{appendix}

\end{document}